# Li$_x$(C$_5$H$_5$N)$_y$Fe$_{2-z}$Se$_2$: a defect resilient expanded-lattice high-temperature superconductor

Alexandros Deltsidis,[a,b] Laura Simonelli,[c] Georgios Vailakis,[a,b] Izar Capel Berdiell,[a] Georgios Kopidakis,[a,b] Anna Krzton-Maziopa,[d] Emil S. Bozin,[e] and Alexandros Lappas[a,*]

[a]*Institute of Electronic Structure and Laser, Foundation for Research and Technology–Hellas, Vassilika Vouton, 71110 Heraklion, Greece*

[b]*Department of Materials Science and Technology, University of Crete, Voutes, 71003 Heraklion, Greece*

[c]*ALBA Synchrotron Light Source, Carrer de la Llum 2-26, 08290 Cerdanyola del Vallés, Spain*

[d]*Warsaw University of Technology, Faculty of Chemistry, Noakowskiego St. 3, 00-664, Warsaw, Poland*

[e]*Condensed Matter Physics and Materials Science Department, Brookhaven National Laboratory, Upton, NY 1197, USA*

[*] e-mail: lappas@iesl.forth.gr

# ABSTRACT

Two-dimensional iron-chalcogenide intercalates display a remarkable correlation of the interlayer spacing with the enhancement of the superconducting critical temperature ($T_c$). In this work, synchrotron x-ray absorption (XAS; at Fe and Se K edges) and emission (XES) spectroscopies, allow to discuss how the important rise of $T_c$ (~ 44 K) in the molecule intercalated $Li_x(C_5H_5N)_yFe_{2-z}Se_2$ relates to the electronic and local structure changes felt by the inorganic host upon doping (x). XES shows that widely-separated layers of edge-sharing $FeSe_4$ tetrahedra, carry low-spin moieties with a local Fe magnetic moment slightly reduced compared to the parent $\beta$-$Fe_{2-z}Se_2$. Pre-edge XAS advises on the progressively reduced mixing of metal $3d$ – $4p$ states upon lithiation. Doping-mediated local lattice modifications, probed by conventional $T_c$-optimization measures (cf. anion height and $FeSe_4$ tetrahedra regularity), become less relevant when layers are spaced far away. On the basis of extended x-ray absorption fine structure, such distortions are compensated by a softer Fe-network that relates to Fe-site vacancies, alleviating electron-lattice correlations and superconductivity. Density functional theory (DFT) guided modification of isolated $Fe_{2-z}Se_2$ (z, vacant sites) planes, resembling the host layers, identify that Fe-site deficiency occurs at low energy cost, giving rise to stretched Fe-sheets, in accord with experiments. The robust high-$T_c$ in $Li_x(C_5H_5N)_yFe_{2-z}Se_2$, arises from the interplay of electron donating spacers and the iron-selenide layer's tolerance to defect chemistry, a tool to favorably tune its Fermi surface properties.

# I. INTRODUCTION

The first iron-based compound that exhibited superconducting properties, below the very modest transition temperature of ~4 K, was the LaOFeP [1] system. Soon after, a major leap forward was demonstrated upon fluorine doping at the oxygen site, with the La[$O_{1-x}F_x$]FeAs[2] derivative, exhibiting a high critical transition temperature ($T_c$) of ~26 K. This caught the scientific community by surprise since it was believed that the large magnetic moment of Fe is sufficiently strong to perturb the pairing of electrons responsible for superconductivity. In the same year, the discovery of β-FeSe, a layered compound with the simple PbO-type structure and $T_c$ of 8 K, made available new exciting horizons[3]. Like the high-$T_c$ cuprates, the iron-based systems seem to exhibit an unconventional superconducting mechanism[4,5], thus steering further the interest of the community. Since then, there was a surge of research activities with the aim to possibly discover even higher-$T_c$ Fe-based compounds, but also to understand the physical mechanisms behind superconductivity.

The quest for high-$T_c$ Fe-based systems was extensively explored by implementing chemical tuning in the structurally simple (11-type) β-FeSe. The latter adopts a tetragonal (P4/nmm) crystal structure at room temperature and is a quasi-two dimensional material composed of slabs of $FeSe_4$ edge-sharing tetrahedra[6]. A dramatic increase of $T_c$ from 8 up to 37 K is observed under hydrostatic pressure[7,8]. Moreover, the van der Waals gap between the FeSe slabs makes it an ideal candidate for intercalation chemistry. Indeed, FeSe layers have been successfully intercalated by alkalis[9–11] or even alkaline earth metals[12], eventually exhibiting $T_c$s as high as 46 K.

Except simple metal cations, solvates of selected alkali or alkaline earth metals with bases (cf. ammonia, aliphatic amines, diamines and aromatic amines) comprise another type of intercalating species that can enter the structure of β-FeSe as electron donors that promote further the $T_c$ enhancement.[13] Interestingly, in these type of Fe-chalcogenide intercalates, there is a correlation of the interlayer spacing (d, defined as the Fe-sheet distance in adjacent FeSe layers, along the c-axis) with the $T_c$, which empirically shows a progressive growth of $T_c$ with increasing d that saturates at a separation of d ~ 8.6 Å[14]. Note that a similar correlation between interlayer spacing and $T_c$ is also exhibited in the case of $Li_xM_yHfNCl$ (M: Molecule) superconductors[15]. The dependence of $T_c$ with interlayer distance was interpreted in view of the weakening of interactions involving successive layers that in effect seem to suppress factors that compete with superconductivity. The latter is demonstrated in a theoretical work where it is suggested that when the Fe-interlayer distance rests around the saturated value (d ~ 8.6 Å), the Fermi surface is nearly two-dimensional [16].

However, tuning the electronic structure of such intercalates appears not to be a bare function of the interlayer distance, but may crucially depend on how efficiently carriers can be injected in the two-dimensional (2D) layers, as it has been demonstrated with the remarkable enhancement of $T_c$ (65 K) in single-layer FeSe films[17,18]. Therefore, optimizing doping[19–22] of FeSe sheets, and controlling its impact on local structural parameters, such as the Fe-Fe distance and the Fe-Se bond length[19], offers another avenue to tune the electronic structure and promote superconductivity. Intended chemical substitutions, as depicted through the phase diagram of the $K_xFe_{2-y}Se_{2-z}S_z$ [23], modify the regularity of the Fe$Ch_4$ (*Ch*= Se, S) key building blocks, believed to be a crucial factor governing the competition of magnetism and superconductivity, having particular influence on electron pairing. The emerging local lattice modification, probed through the anion height (i.e. Se to Fe-sheet c-axis normal distance, $h_z$)[24], is believed to govern the relative positions of the Fe 3*d* bands and mediate the evolution of the superconducting critical temperature[25]. Equally significant is the sensitivity of $T_c$ to the in-plane Fe-site vacancy concertation as it is demonstrated by the expanded lattice $Li_{1-x}Fe_x(OH)Fe_{1-y}Se$ (d~ 9.3 Å), Fe-deficient system[21]. In the latter, a mere ~95% Fe-site occupancy leads to a strikingly lower $T_c$ (~17 K), as compared to that in the vacancy-free derivative (~41 K), highlighting the importance of the local bonding environments. Indeed, defects may be crucial for type-II superconductors as they can act as pinning centers of the vortices carrying the magnetic flux, thus increasing the current carrying capability ($J_c$), or instead turn into pairing breakers, thus suppressing $T_c$[26,27]. Therefore, deciphering subtle differences in the atomic structure configuration that lead to markedly different physical properties is of fundamental importance in Fe-based correlated electron systems.

Among the various doping methods, electron donor molecules co-intercalated with alkali metals in the β-FeSe set the stage to study the impact on the electronic structure of the host, as the intercalation of [alkali metal - molecule adducts] not only increases the interlayer separation but also leads to an important rise of the $T_c$. Along this direction, low-temperature, solvothermal routes were employed to afford molecule-intercalated compounds, with large Fe-interlayer distance (d > 8.6 Å), at a formal composition of $Li_x(C_5H_5N)_yFe_{2-z}Se_2$ (PyH5 = $C_5H_5N$; Pyridine)[28]. Challenges in their growth have been recently addressed by high-throughput, in-situ synchrotron x-ray total scattering studies.[29] Insights are now sought on the way interlayer guests [Li-PyH5], impact (a) the evolution of the FeSe$_4$ building blocks, in the FeSe electronically active layers and (b) the magnitude of $T_c$. For this, element-specific, synchrotron x-ray absorption (XAS, at the Fe and Se K-edges) and emission (XES) spectroscopies have been chosen to uncover the local structure correlations with the evolving electronic properties, including the Fe local magnetic moments, as a function of doping across the $T_c$ (~44 K).

While the near edge features accessed the unoccupied electronic states, the extended x-ray absorption fine structure (EXAFS) provided quantitative information on the FeSe layer correlated local distortions. The work discusses how the square planar Fe-Fe network, accompanied by 'taller' than optimal ($h_z$~1.38)[24] anion heights, is systematically stretched in $Li_x(C_5H_5N)_yFe_{2-z}Se_2$. Analyses offer evidence for a relaxed, softer sublattice, proposing a depleted Fe-site occupancy upon intercalation. Its compensating role makes conventional $T_c$ optimization measures (cf. $h_z$) less relevant when layers are spaced far away. Experiments are complemented by DFT calculations on defected $Fe_{2-z}Se_2$ (z= 0, 0.08, 0.22) monolayers mimicking the strong separation of the inorganic slabs in the intercalants. Theory postulates a favorable Fe-vacancy formation in isolated, 2D inorganic layers, with concomitant in-plane lattice expansion, as seen by EXAFS. Arguments are offered in view of the defect-mediated electronic structure of single-layer FeSe and its relation to the robust superconducting state of the titled, expanded lattice, molecule intercalated Fe-selenides.

## II. METHODS

### A. Experiments

Polycrystalline β-FeSe parent phase was prepared by employing established solid-state synthesis protocols[30]. Powder samples of its intercalated derivative, $Li_x(PyH5)_yFe_{2-z}Se_2$, with nominal composition x= 0.2, 0.6 and 1.0, were synthesized by means of a modest temperature (80°C), solvothermal method[29]. Sample manipulations were undertaken inside in an Ar-circulating MBRAUN (UNILab) glovebox, with <1 ppm $O_2$ and $H_2O$. Basic characterization of the materials was performed by powder X-Ray diffraction (PXRD), thermogravimetric analysis (TGA), and AC susceptibility (SI §S1). Phase purity was tested by PXRD, using a Bruker D8 Advance diffractometer, with Cu-$K_a$ radiation (Fig. 1a). Intense (00$l$) reflections, indexed on the basis of the $ThCr_2Si_2$ structural type, confirmed widely-separated FeSe layers (d~ 16.2 Å), suggesting successful intercalation of the β-FeSe host (d~ 5.5 Å) by [Li-PyH5] moieties. Thermal analysis was performed with a TA Instruments SDT Q600 simultaneous TGA / DSC system. From the weight loss curves of TGA (Fig. S1b; SI §S1), the average content of the molecular intercalant per formula unit of $Li_x(C_5H_5N)_yFe_{2-z}Se_2$ was found to vary with the Li concentration and estimated as $0.30 \leq y \leq 0.49$ (Table S1). AC susceptibility measurements ($H_{ac}$= 1 Oe and f= 999 Hz) were performed with an Oxford Instruments Maglab EXA 2000 vibrating sample magnetometer (Fig. 1b,c). Normalized data ($4\pi\chi = -1$; SI units) revealed the five-fold rise of $T_c$ in the expanded lattice materials ($T_c$~ 44 K), compared to the parent phase ($T_c$~ 8 K).

Fe ($E_0$ = 7112 eV) and Se ($E_0$ = 12658 eV) K-edge XAS measurements were performed in transmission mode at the beamline BL22-CLÆSS of the ALBA synchrotron,[31] where x-rays emitted by a multipole wiggler were monochromatized using a Si(311) double crystal monochromator, with Rh-coated mirrors used to reject higher harmonics. Samples in powder form were mixed with BN in order to optimize the absorption jump in the two edges. The mixed powders were pressed into pellets of 5 mm diameter. Samples were mounted on a continuous flow He cryostat in which the temperature was maintained within ±1 K through the 20-300 K range. Several scans were collected for any given temperature to ensure reproducibility and improve the signal-to-noise ratio. Data normalization and modelling (SI §S1) was performed with the DEMETER suite of programs[32].

The XES spectra were collected in back scattering horizontal geometry by means of the CLEAR emission spectrometer. The spectrometer is based on a diced Si(333) dynamically bent analyzer crystal (bending radius R = 1 m) and a 1D position-sensitive Mythen detector. The Fe K$\beta$ emission lines were measured by exciting the sample well above the Fe K-edge, with a total energy resolution around 0.8 eV. XES measurements were performed on pellets at 300 and 20 K only, with several scans taken to ensure spectral reproducibility and high signal-to-noise-ratio. The data were normalized to $I_{max}$ of the K$\beta_{1,3}$ emission line and then divided by the integrated intensity of each spectrum in the region of 7028-7078 eV, with a subsequent linear background subtraction to bring their baseline to zero.

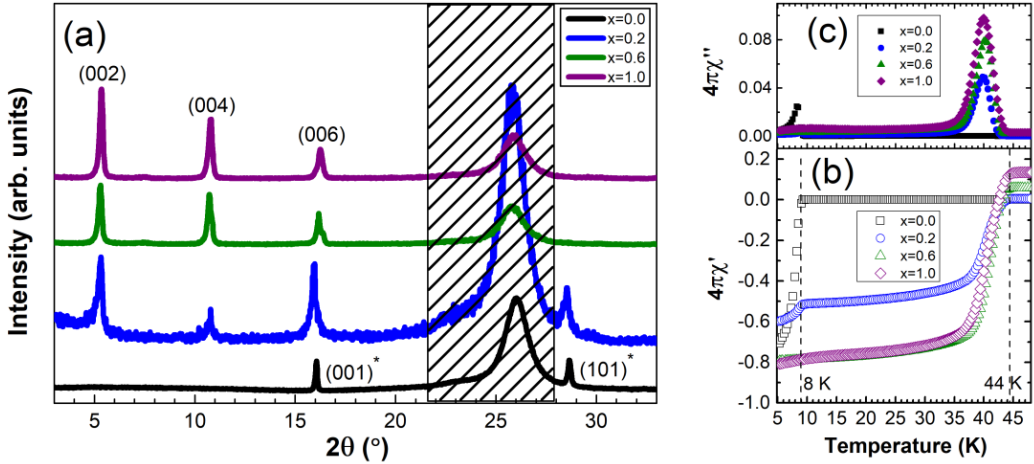

**Figure 1.** (a) powder XRDs (Cu-K$_a$) of the corresponding Li$_x$(C$_5$H$_5$N)$_y$Fe$_{2-z}$Se$_2$ (x = 0.0, 0.2, 0.6 and 1.0) specimen. Shaded region of XRD corresponding to 2$\theta$ = 20-28° is excluded due to contribution from the home-made, air-tight sample holder. The Bragg reflections of the intercalated phase are indexed on the basis of the ThCr$_2$Si$_2$ (122) structural type.[29] The (*hkl*)$^*$ indices, denote reflections belonging to the PbFCl (11) structural type of the parent β-FeSe. AC susceptibility (H$_{ac}$ = 1 Oe, f = 999 Hz) for the Li$_x$(C$_5$H$_5$N)$_y$Fe$_{2-z}$Se$_2$ (x = 0.0, 0.2, 0.6 and 1.0) samples, where (b) is the real part ($\chi'$) and (c) the imaginary part ($\chi''$) of the susceptibility. The volume susceptibilities have been normalized so that 4$\pi\chi$ = -1 (CGS).

## B. Theoretical calculations

In order to study the effect of Fe vacancies on the atomic and electronic structure of isolated FeSe layers, relevant to the expanded lattice Li$_x$(C$_5$H$_5$N)$_y$Fe$_{2-z}$Se$_2$ system, first principles calculations were performed within DFT using the Vienna Ab initio Simulation Package (VASP) [33,34]. Projector augmented wave method (PAW) for core electrons and nuclei[35] and the generalized gradient approximation (GGA) of Perdew-Burke-Ernzerhof (PBE) for the exchange-correlation functional[36], were employed. Wave functions were expanded on a plane wave basis set with a kinetic energy cut-off of 450 eV. The k-sampling of the 1$^{st}$ Brillouin zone (BZ) was selected based on a Monkhorst-Pack grid of 17 × 17 × 1 for the primitive unit cell and was appropriately scaled for the designed supercells in a way that the sampling is similar in all systems. Lattice vectors and the atomic positions were fully relaxed, until the norms of all atomic forces were smaller than $5 \cdot 10^{-3}$ eV/Å, with the electronic self-consistent loop break condition set at $10^{-7}$ eV, which became $10^{-8}$ eV for the electronic band structure calculations. Hence, spin-polarized DFT results are presented for the single-layer pristine $Fe_2Se_2$ and its defected $Fe_{2-z}Se_2$ derivatives ($z$, the stoichiometric concentration of Fe vacancies), mimicking the isolated iron-selenide slabs of $Li_x(C_5H_5N)_yFe_{2-z}Se_2$. For the $Fe_{2-z}Se_2$ models, it was necessary to build supercells of different sizes in order to study the properties upon the varying degree of Fe vacancy ($z$= 0.08, 0.22) in the two-dimensional layers. The corresponding effective band structures were accessed by unfolding (SI §S1) from the supercell to the parent $Fe_2Se_2$ unit cell BZ. Periodic boundary conditions were applied, and in order to avoid interaction between periodic images, a vacuum of at least 20 Å in the out of plane direction was added. The effect of electron doping was taken into account by assigning excess electrons on the empty states above the Fermi level ($E_F$).

## III. RESULTS AND DISCUSSION

### A. Electronic structure insights

#### 1. Absorption spectra

The x-ray absorption near edge structure (XANES) region is rich in information regarding the electronic state of the material. Representative normalized Fe and Se K-edge spectra, collected at 20 K and 300 K (RT) are shown in Fig. 2a-b. The 20 K dataset is selected since the intercalated phase is in its superconducting state (Fig. 1b), while the RT spectra represent the normal state. With increasing Li content ($0 \leq x \leq 1.0$) changes in the near-edge features are observed.

The Fe K-edge XANES reveal (Fig. 2a) similar absorption features to those reported for the Fe-based pnictides[37,38], as well as other Fe-based chalcogenide superconductors[39,40]. The comparison with spectra of a known $Fe^{2+}$ standard[37,38] confirm a near 2+ formal valence state in the investigated samples. The Fe K-edge (Fig. 2a) is governed mainly by the $1s \rightarrow \varepsilon p$ dipolar transition[39–41]. The characteristic pre-edge peak (#A) at low energy, becomes accessible due to lack of inversion symmetry in the $FeSe_4$ tetrahedral units, which enables mixing of the metal $3d$ and $4p$ orbitals, giving rise to a combination of $1s \rightarrow 3d$ quadrupole and stronger $1s \rightarrow 4p$ dipole transitions.[42] It is sensitive to the spin state, oxidation state and local geometry[43], while its intensity can be related to the level of the orbital hybridization[44]. Feature #B, which appears as a shoulder in the absorption jump (Fig. 2a) arises from Fe $1s \rightarrow 4p$ transition [39,45], while feature #C at 1.5 eV above $E_0$ corresponds to the $1s \rightarrow 4p$ admixed with the Se $d$ transition [39,45]. Features at higher energies are mainly due to the photoelectron multiple scattering with the nearest neighbors.

When electronic doping of the FeSe matrix takes place, this affects the resolved components (#A, #B, #C). Indeed, a Gaussian fit (SI §S2) to the pre-edge peak (#A) for samples measured at 20 K, reveals the intensity reduction from the parent (β-FeSe) to the expanded lattice materials (Fig. S2). The diminution of intensity at #A by intercalation (inset, Fig. 2a), implies a modest decrease in the Fe $3d$ and Fe $4p$ orbitals mixing, due to weakening of non-centrosymmetry at the Fe-sites [cf. evolution of Se-Fe-Se bond angle (α) and Fe-Se bond length; *vide-infra*] [40,44]. Similarly, feature #B, is reduced with intercalation, advising on the subtle evolution of the Fe sites coordination symmetry and its sensitivity to shakedown transitions due to Se to Fe charge-transfer effects. Feature #C systematically decreases in intensity by doping (inset, Fig. 2a), in agreement with the progressive filling of the corresponding band. Interestingly, feature #C decreases in intensity by cooling down to 20 K, for all the systems, including the parent β-FeSe. Since the latter is not yet in the superconducting state, the effect most likely corresponds to a structural thermal contraction.

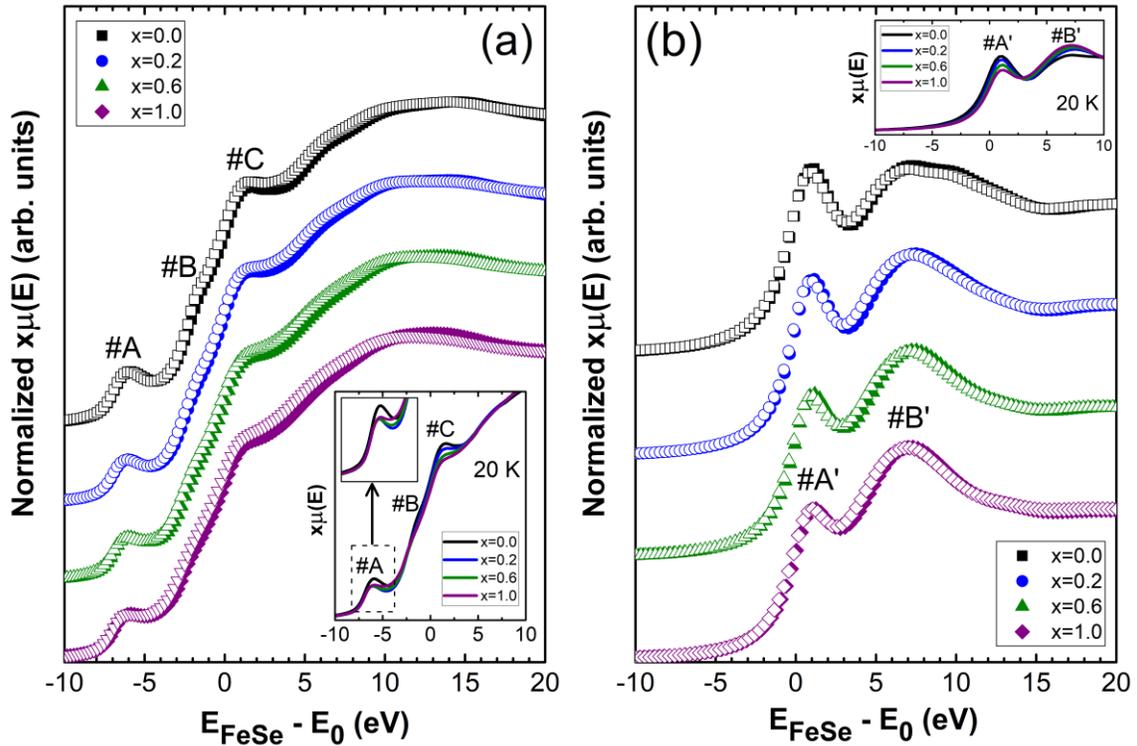

**Figure 2.** X-ray absorption near-edge structure versus energy for the parent compound β-FeSe (x = 0.0) and a series of molecule intercalated iron selenides $Li_x(C_5H_5N)_yFe_{2-z}Se_2$ (x=0.2, 0.6, 1.0) at 20 K (filled symbols) and at 300 K (open symbols), measured at (a) the Fe K-edge and (b) the Se K-edge. Labels, #A (A'), #B (B') and #C, mark near-edge spectral features across the samples of interest. Data are shifted vertically for clarity. Insets: The absorption lines over-plotted and zoomed, revealing subtle intensity variations. The x-axis is presented after subtracting the $E_0$ for each element, from the incoming energy, E(k).

The corresponding, normalized Se K-edge spectra (Fig. 2b), reveal typical features of $Se^{2-}$ systems[39,46,47]. They consist mainly of two features: (i) the sharp peak #A' at 1 eV above the absorption edge due to the direct Se $1s \rightarrow 4p$ dipole transition[39], and (ii) the much broader peak/feature #B' (~7 eV above the edge), which is governed by the multiple scattering of the photoelectron with its neighbors, thus it is closely related to the local atomic arrangement[39,40]. Feature #A' represents the fine structure near the core edge and its intensity is related to the unoccupied density of states near $E_F$. It is evident that doping has a large effect on both peaks #A' and #B'. Peak #A' progressively decreases in intensity, suggesting a decreased number of unoccupied Se $4p$ states near $E_F$ in the doped systems (inset, Fig. 2b). Peak #B' shows the opposite trend, where the doping increase leads to a more intense peak, indicating a possible reorganization of the lattice.

## 2. X-ray Emission spectra

With the purpose to study the relationship between superconductivity and the Fe magnetic properties, Fe K$\beta$ XES were measured at 20 K (T< $T_c$) and at 300 K (T> $T_c$). The fluorescent emission expressed through the K$\beta$ lines corresponds to the $3p \rightarrow 1s$ transitions[48]. The intra-atomic interaction between the Fe $3p$ and $3d$ unpaired electrons is the dominant mechanism that shapes the K$\beta$ spectra[48,49]. The $3p$-$3d$ exchange interactions split the K$\beta$ spectrum into a strong K$\beta_{1,3}$ peak and a broad K$\beta'$ shoulder at lower energies. The energy separation between the K$\beta_{1,3}$ and K$\beta'$ is approximately given by $\Delta E = J(2S + 1)$, where $J$ is the exchange integral between the electrons in the $3p$ and $3d$ shells and $S$ is the net spin of the unpaired $3d$ electrons[50]. This establishes the K$\beta$ emission lines as an indirect spin probe. In the general case, the K$\beta_{1,3}$ peak position may shift with the net spin in a linear fashion, while the absence of a K$\beta'$ component advises on a low-spin (LS) compound [cf. presence of K$\beta'$ points to a high-spin (HS) case] [48,49]. Care should be taken though, when interpreting the K$\beta$ spectra, as charge-transfer (covalency) processes may also (a) decrease the splitting of the K$\beta_{1,3}$ and K$\beta'$ features, and (b) increase the intensity of the K$\beta_{1,3}$, while suppressing the K$\beta'$.[49] As such observations are not altogether resolved at the level of the present data, to a first approximation, we approach the K$\beta$ lines as a local spin probe. Figure 3a presents the Fe K$\beta$ XES for samples with varying Li content ($0.2 \leq x \leq 1.0$) and compares them to the parent β-FeSe (x = 0). The Fe K$\beta$ emission lines show a negligible change with temperature or doping. While the weak K$\beta'$ feature, throughout the series (Fig. 3a), offers a subtle indication of a low valence spin for Fe, [48] closer examination of K$\beta_{1,3}$, reveals a small change from β-FeSe to the expanded lattice $Li_x(C_5H_5N)_yFe_{2-z}Se_2$ materials (inset, Fig. 3a), albeit very small. These variations are seen in the difference (Fig. 3b), derived by subtracting the 300 K spectra of β-FeSe from those of the doped compounds. They can be quantified by utilizing the integrated absolute difference spectra (IAD; SI §S2) method[51], enabling the extraction of the respective IAD values (inset, Fig. 3b). Since the IAD is calculated with respect to the system with the highest Fe local magnetic moment, the raised IAD values indicate a slightly lower Fe local magnetic moment in the expanded lattices when compared to that in β-FeSe ($2\mu_B$)[41,52], an effect also consistent with the excess electrons of the intercalated materials.

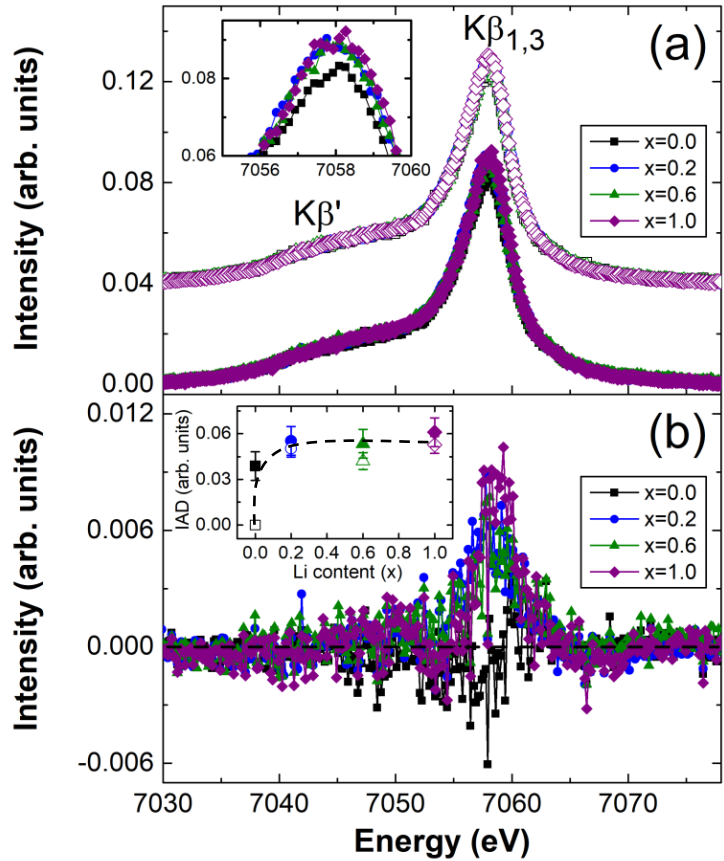

**Figure 3.** X-ray emission spectra $Li_x(C_5H_5N)_yFe_{2-z}Se_2$ ($0 \leq x \leq 1$): (a) Fe K$\beta$ spectra collected at 20 K (filled symbols) and at 300 K (open symbols); data are shifted vertically for clarity. Inset: zoom around the K$\beta_{1,3}$ peak, showing a subtle change in the intensity at 20 K. (b) K$\beta$ emission spectra differences at 20 K and the integrated absolute difference (IAD; inset) revealing the local spin moment variation with respect to the parent β-FeSe.

### B. Local structure insights

While the XANES is limited to a few eVs around the edge, insights on the local lattice structure modifications are provided by the EXAFS region, starting few tens of eV beyond the edge. Here sinusoidal oscillations of the signal are viewed in the reciprocal space (Fig. S3), with the $k^2$-weigthed EXAFS oscillations extracted from the Fe and Se K-edge XAS. Measured between $20 \leq T \leq 300$ K, the signal is visible up to a high wave vector (k) (Fig. S3), with evolutions suggesting: (i) damping of the signal with increasing T and doping, and (ii) subtle structural changes.

The Fourier transform (FT) of the Fe and Se K-edge EXAFS oscillations at representative temperatures (20 and 300 K) are reported in Fig. 4a,b. The FTs of the $k^2$-weighted EXAFS are performed in the k-range of 3.12-15 Å$^{-1}$ and 3.55-15 Å$^{-1}$ for the Fe and Se K-edge spectra, respectively. A Hanning window in the R range of 1.5 - 3 Å has been utilized for fitting the data. The FTs of the EXAFS due the nearest-neighbor pairs are dominated by contributions from the Fe-Se (~2 Å) and Fe-Fe (~2.4 Å) distances (Fig. 4a) in the Fe K-edge, while for the Se K-edge solely by the Se-Fe bond distances (Fig. 4b). It should be noted that the actual bond distances are longer than reported in Fig. 4, as the FTs were not corrected for the phase-shift implemented in the backscattered wave. Peaks at longer R (> 3 Å) are due to single scattering contributions of distant coordination shells and multiple scatterings involving different paths. Quantitative information on the local structure evolution upon doping is extracted by fitting a structural model to the FT of the EXAFS signal in the context of single scattering approximation (eq. 1; SI §S1). The starting crystal model utilized in the analysis discussed here, is the simple tetragonal (P4/nmm) structure of the layered β-FeSe [30].

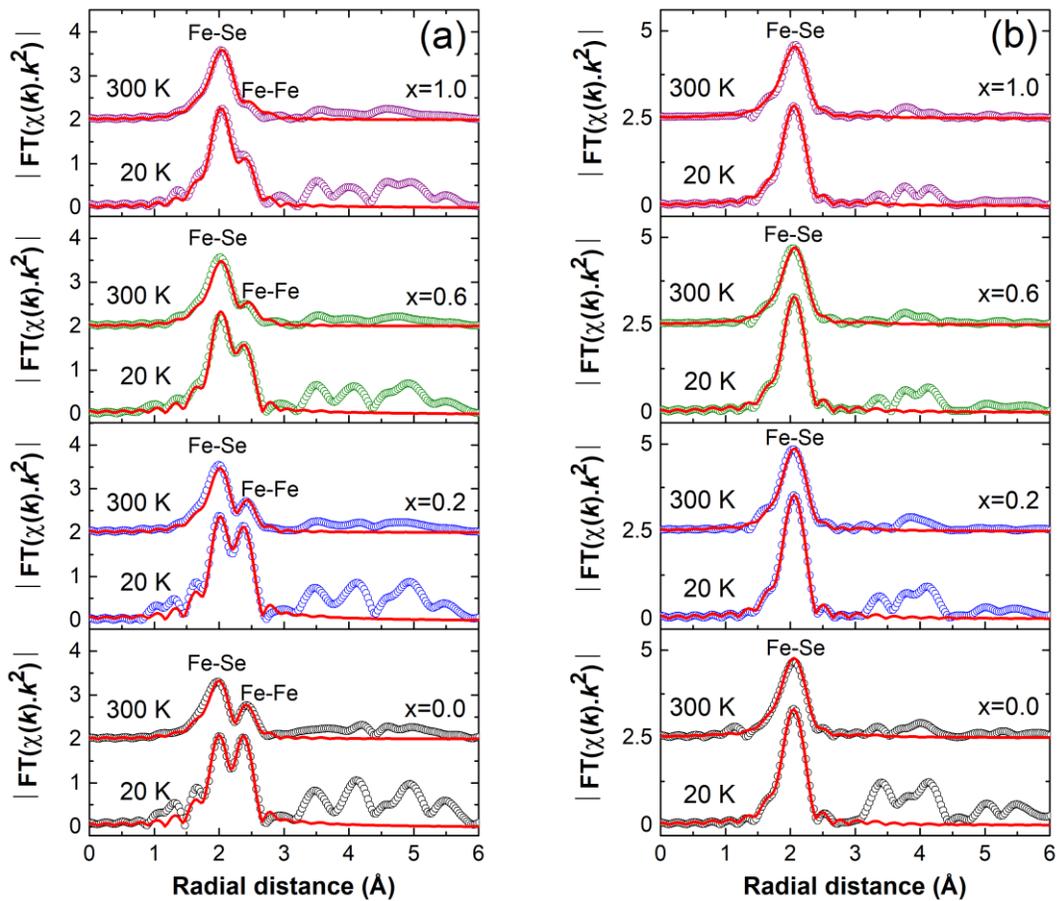

**Figure 4.** Temperature evolution of the Fourier transform (FT) of the EXAFS oscillations in $Li_x(C_5H_5N)_yFe_{2-z}Se_2$ ($0 \leq x \leq 1$) measured at the Fe K-edge (a) and the Se K-edge (b); the FTs are not corrected for the phase shifts and represent raw experimental data. The optimal R-range, 1.5-2.75 Å, is chosen for least square fitting of the local lattice model to the data; data and fits are shown with open circles and solid lines, respectively.

## 1. Intercalant-driven local distortions

Since the discovery of superconductivity in Fe-chalcogenides, the correlated local distortions within FeSe layers were claimed to be important for tuning the materials' electronic structure[19]. Therefore, the present analysis is performed in the R range of 1.5 - 3 Å, describing primarily the local bonding deformations within one layer of the β-FeSe structure type, incorporating the square planar Fe-Fe network in the ab plane of the unit cell, and the Se capping layers above and below the basal plane (Fig. 5).

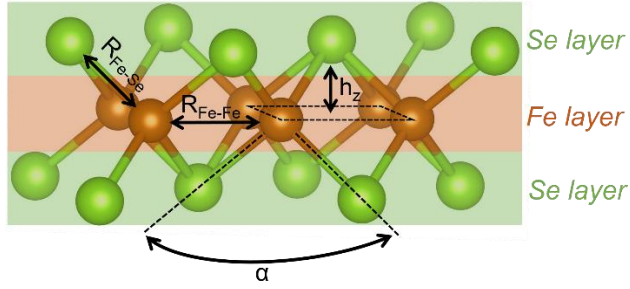

**Figure 5.** Schematic depicting the local lattice environments, involving the in-plane FeSe$_4$ edge-sharing tetrahedral units, defined by the bond lengths ($R_{Fe-Se}$ and $R_{Fe-Fe}$), the anion height ($h_z$) and Se-Fe-Se bond angle (α).

Therefore, from the analysis of Fe K-edge EXAFS of Li$_x$(C$_5$H$_5$N)$_y$Fe$_{2-z}$Se$_2$, the local Fe-Fe and Fe-Se bond distances were obtained. When the amount of doping is increased, the Fe-Fe distance becomes systematically longer ($\Delta R_{Fe-Fe}$ ~ 0.02 Å; Fig. 6a), while that for the Fe-Se shows less substantial lengthening ($\Delta R_{Fe-Se}$ ~ 0.01 Å; Fig. 6b). With respect to tetragonal β-FeSe, there is a discernible expansion of ~1% in the basal plane lattice parameter (a=b= $\sqrt{2}$ ×Fe-Fe) at ambient temperatures. Moreover, the bond evolution upon intercalation reflects also in a subtle widening of the Se-Fe-Se bond angle (α; Fig. 5) within the FeSe$_4$ tetrahedral coordination (SI §S3; Fig. S4a), which however does not deviate much from the distorted tetrahedral geometry (α ~ 104.5°) observed in the pristine β-FeSe[53,54]. The evolution of the bond angle reflects the relaxation of the FeSe layer once the Fe-Se and Fe-Fe bonds are stretched when electron-donating species [Li-PyH5] are intercalated in the FeSe matrix. In addition, implies that the tetrahedral regularity (α ~ 109.47°)[55] of the FeSe$_4$ units is not necessarily a required condition for optimizing superconductivity. The above support subtle modifications in the non-centrosymmetry of the Fe local coordination with lithiation that are also consistent with the lowering of the intensity in the XAS pre-edge structure (Fig. 2a).

Although the elongation of the intralayer Fe-Fe/Se bonds may indicate electron donation effects, assessing the anion height ($h_z$; Fig. 5), offers a key layer metric (SI §S3)[24] that is known to mediate the materials' Fermi surface topology[25,56]. For the Li$_x$(C$_5$H$_5$N)$_y$Fe$_{2-z}$Se$_2$ series, the anion height is $h_z$ ~ 1.47 Å (T = 300 K, Table 1; Fig. S4b), and appears insensitive to the Li level. However, as the parent β-FeSe$_{1-x}$[57,58] and intercalated materials display a similar $h_z$ (~ 1.47 Å), the five-fold enhancement of the T$_c$ in the $Li_x(C_5H_5N)_yFe_{2-z}Se_2$ cannot be merely due the adjustment of the Fermi surface by $h_z$.[25] Other $T_c$-scaling parameters, such as the raising separation of the Fe-sheets from d ~5.5 to ~16.5 Å and the electron donating capability of the molecular spacers intercalated in the interlamellar space, may come into play. It is worth noting that a comparable local lattice distortion ($h_z$ ~ 1.46 Å) is also met in the related $Li_x(NH_2)_y(NH_3)_yFe_{2-z}Se_z$ (d~ 8.4 Å, $T_c$~ 43 K)[59] and $Li_{1-x}Fe_x(OH)Fe_{1-y}Se$ (d~ 9 Å, $T_c$~ 41 K) [21] compounds, corroborating that intercalants may promote doping of the FeSe layers and the raised $T_c$. It appears that priming the layers far apart in the $ThCr_2Si_2$-type phases by intercalating large molecule-based donors, the effects of $h_z$ and $FeSe_4$ tetrahedral regularity become less important for optimising $T_c$ further.

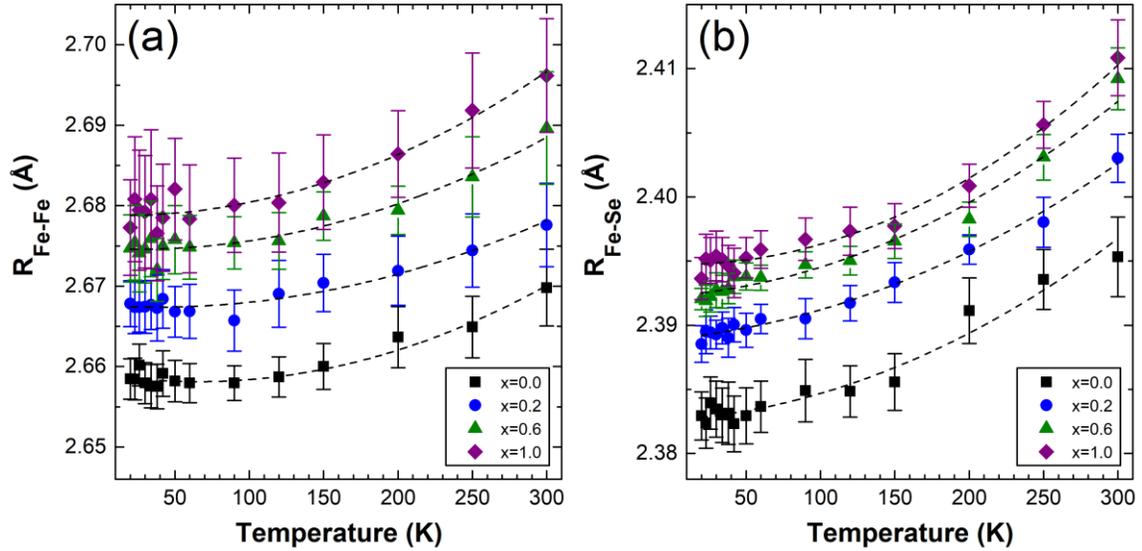

**Figure 6.** Temperature evolution of the bond correlations in the $Li_x(C_5H_5N)_yFe_{2-z}Se_2$ ($0 \leq x \leq 1$) series: (a) the Fe-Fe pairs of distances, extracted from the Fe K-edge, and (b) the Fe-Se pairs of bonds, extracted from the Se K-edge. The dashed lines are a guides to the eye.

Comparisons against related systems may reveal useful views about intralayer atomic correlations relevant to properties. For example, in the $K_{0.8}Fe_{1.6+x}Se_2$, with the $ThCr_2Si_2$ structure type ($T_c$ up to ~30 K), although the Fe-Se bond (~2.41-2.42 Å)[60,61], compares well with that in the highly lithiated $Li_{1.0}(C_5H_5N)_yFe_{2-z}Se_2$ intercalate, its Fe-Fe distance (~2.74 Å) [60] is significantly longer. Interestingly, earlier EXAFS on $K_{0.8}Fe_{1.6+x}Se_2$, conferred structural evidence that key factors affecting the Fe-Fe local bonding were (i) the largely disordered structure [62,63,64] and (ii) the reduced Fe-site occupancy. Both, heavily impact the Fe-site contribution to the FT of the EXAFS signal in Fe-deficient selenide derivatives [21,60]. Indeed, such a systematic loss of signal in the Fe-Fe feature of the FT data (cf. R ≅ 2.4 Å, Fig. 4a) is also resolved in the $Li_x(C_5H_5N)_yFe_{2-z}Se_2$ series with increased Li content, thus implying increased concentration of Fe vacancies.

**Table 1.** Comparison of parameters extracted from the least square fits of the $Li_x(C_5H_5N)_yFe_{2-z}Se_2$ EXAFS data (see text). The superconducting (SC) volume fraction is extracted from the AC susceptibility measurements. The values for bond lengths ($R_{Fe-Fe}$, $R_{Fe-Se}$), anion-height ($h_z$) reported here are the respective values at 300 K. Einstein temperature ($\theta_E$), MSRDs for the static disorder ($\sigma_0^2$) and Fe-Fe lattice force constant ($k$) were determined from the corresponding least square fits based on the correlated Einstein model (see text). The coordination number (N) was refined by fitting the 20 K Se K-edge data, i.e., the number of nearest neighbors from the Se site is reported.

| Nominal composition | SC vol. fraction | $R_{Fe-Fe}$ (Å) | $R_{Fe-Se}$ (Å) | $h_z$ (Å) | $\theta_E^{Fe-Fe}$ (K) | $\theta_E^{Fe-Se}$ (K) | $\sigma_0^{Fe-Fe} \times 10^{-4}$ (Å²) | $\sigma_0^{Fe-Se} \times 10^{-4}$ (Å²) | $k^{Fe-Fe}$ (eV Å$^{-2}$) | N |
|---|---|---|---|---|---|---|---|---|---|---|
| x = 0.0 | 70% | 2.669(4) | 2.395(3) | 1.474(9) | 263.3 ± 5.3 | 318.2 ± 4.8 | 0.0 ± 0.9 | 2.1 ± 0.5 | 3.4 ± 0.1 | 3.97 ± 0.11 |
| x = 0.2 | 60% | 2.677(5) | 2.403(2) | 1.479(8) | 261.5 ± 3.3 | 319.7 ± 2.7 | 0.0 ± 0.9 | 0.0 ± 0.4 | 3.4 ± 0.1 | 4.05 ± 0.08 |
| x = 0.6 | 75% | 2.689(7) | 2.409(2) | 1.478(9) | 258.2 ± 3.6 | 319.6 ± 5.6 | 6.1 ± 0.9 | 1.5 ± 0.6 | 3.3 ± 0.1 | 3.99 ± 0.06 |
| x = 1.0 | 67% | 2.696(7) | 2.411(3) | 1.475(9) | 235.2 ± 3.3 | 319.4 ± 5.0 | 32.5 ± 1.1 | 10.3 ± 0.5 | 2.7 ± 0.1 | 3.88 ± 0.09 |

## 2. Correlated lattice disorder

As effective changes in the local bonding tend to be accompanied by variations in the atom distance-distance correlations within the lattice, we examined their Mean Square Relative Displacements (MSRDs; Debye-Waller factors). The MSRDs for a pair of atoms reflect the lattice configurational disorder expressed through the sum of temperature-independent ($\sigma_0^2$; static) and temperature-dependent ($\sigma_d^2$; dynamic) terms, $\sigma^2 = \sigma_0^2 + \sigma_d^2(T)$ [65]. The evolution of $\sigma_d^2$ for the Fe-Fe distances, for every $Li_x(C_5H_5N)_yFe_{2-z}Se_2$ ($0 \leq x \leq 1$) sample, reveals a systematic increase with Li (Fig. 7a), with no anomaly across the $T_c$. Subsequent analysis on the basis of the Einstein model (eq. 2, SI §S1), suggests a $\theta_E = 263 \pm 5.3$ K for x = 0.0 that is consistent with earlier reports on the binary β-FeSe[66]. However, $\theta_E$ is somewhat reduced for the intercalated compounds (x= 0.2, 0.6; Table 1), implying comparable local Fe-Fe bond correlations at low doping levels. Interestingly, at elevated Li content, $\theta_E = 235.2 \pm 3.3$ K is substantially reduced in $Li_{1.0}(C_5H_5N)_yFe_{2-z}Se_2$, suggesting large changes in these bonds, reminiscent of the $K_{0.8}Fe_{1.6+x}Se_2$ case ($\theta_E \sim 208$ K)[60]. Effectively, the local lattice force constant, $k = \mu\omega_E^2$ (S1 §SI), for the Fe-Fe network is estimated as $k= 2.7\pm0.1$ eV Å$^2$ for the x = 1.0, compared to $k \sim 3.4\pm0.1$ eV Å$^2$ for the others (x= 0, 0.2, 0.6; Table 1). Such a reduction in $k$ upon [Li-PyH5] intercalation suggests that the Fe-Fe bonds become softer at x= 1.0. The relatively relaxed Fe-local environment, is reflected also in the elongated bonds (Fig. 6a) and Se-Fe-Se angles (Fig. S4a). In the case of [Li-NH3] intercalants[59], where $\theta_E \sim 248$ K, a $k \sim 3.1$ eV Å$^{-2}$ for the Fe-Fe network rests between the present samples (Table 1). On the other hand, the extreme changes in the Fe-Fe bond distances, witnessed in the defect-mediated $K_{0.8}Fe_{1.6+x}Se_2$ case[60], result in a much more reduced $k \sim 2.1$ eV Å$^{-2}$. Such comparisons provide useful insights, as the degree of lattice softening can reflect the Fe-sublattice, site-depletion effects.

Electron-lattice correlations are known to be important for superconductivity and MSRDs occasionally express an anomaly when samples cross the $T_c$. For different superconductors, such as $Nb_3Ge$[67], cuprates[68] and iron-based systems, as the [Li-NH3] intercalated derivative[59], MSRDs suggest a local mode hardening. For [Li-PyH5] intercalants though, when warming through the $T_c$ no abrupt changes in $\sigma_d^2(T)$ are detected that would support correlations of electron-lattice interactions and superconductivity (Fig. 7a). While widely separated layers, with bulky molecular spacer in the present case may screen the anomaly at $T_c$, a large FeSe slab occupational disorder may further relax such correlations. The latter is postulated by the substantially larger static disorder ($\sigma_0^2 \sim 0.003$ Å$^{-2}$; Fig. 7a) in the x = 1.0 sample, reminiscent of a similar to $K_{0.8}Fe_{1.6+x}Se_2$ ($\sigma_0^2 \sim 0.0038$ Å$^{-2}$) [60,69] Fe-site disorder.

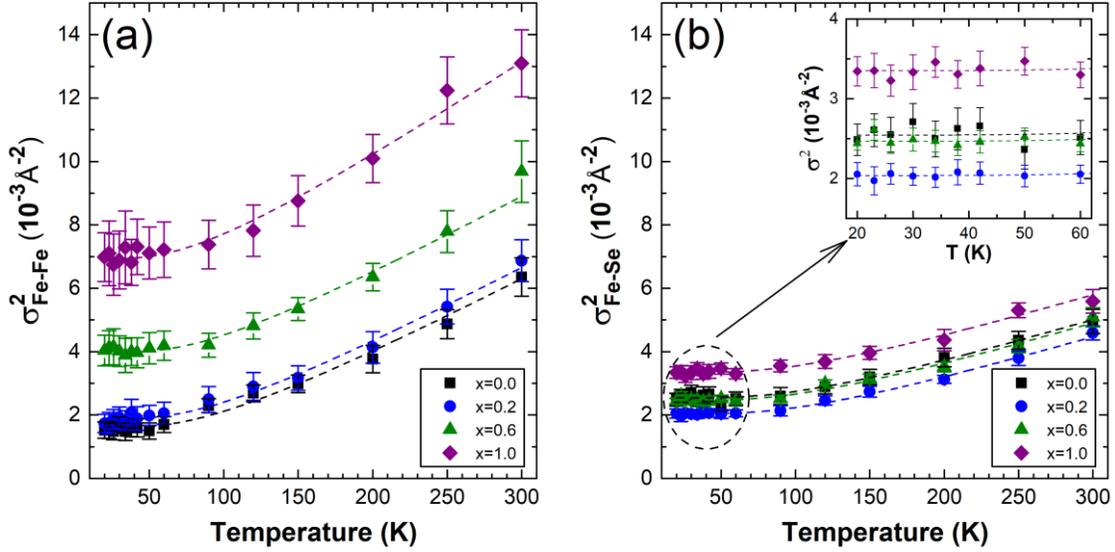

**Figure 7.** Temperature evolution of the MSRDs ($\sigma^2$) for $Li_x(C_5H_5N)_yFe_{2-z}Se_2$ ($0 \leq x \leq 1$) samples: (a) for Fe-Fe pairs extracted from the Fe K-edge, and (b) for Fe-Se pairs extracted from the Se K-edge. Inset: Zoom in the temperature range (20-60) K revealing the subtle variations of the MSRDs. The dashed lines in (a, b) are the corresponding least square fits based on the correlated Einstein model (see text).

When the Fe-Se bond $\sigma_d^2(T)$ evolution is assessed (Fig. 7b), $\theta_E$ for the parent (~318± 4.8 K) and the intercalated derivatives (~319 ± 5K) barely vary, suggesting a similar $k$~ 5.9±0.1 eV Å$^2$. Earlier EXAFS findings agree that the Fe-Se local bond correlations are similar not only to those in β-FeSe, but also across different intercalated derivatives, ranging from short (cf. $Rb_{1-x}Fe_{2-y}Se_2$ [40], $K_{0.8}Fe_{1.6+x}Se_2$ [60,69]) to expanded inter-plane lattice (cf. $Li_x(NH_3)_yFe_{2-z}Se_z$)[59] structures. The Fe-Se bond remains stiff (cf. Fe-Fe), covalent and rather insensitive, even when the layers are severely decoupled as in the case of the [Li-PyH5] intercalates. However, with rising Li content, the magnitude of the $\sigma_0^2$ term (~ 0.001 Å$^{-2}$) grows an order of magnitude for the high Li level (x= 1.0, Table 1) with respect to β-FeSe, adding also support to a rising static disorder in the ab-plane local bonding. Within the limitations of the current analysis, refining the coordination number (N) of nearest-neighbor Fe atoms around the Se absorbing site in the EXAFS model (S1 §SI), a near-stoichiometry in all compositions is suggested, apart from the high Li content derivative (x= 1.0), where N~ 3.88(9), likely points to Fe vacancies (Table 1). This is complemented by the progressive amplified damping of the FT magnitude of EXAFS related to Fe-Fe distances (cf. R ≅ 2.4 Å; Fig. 4a), proposing Fe-site occupational disorder with raised Li-content in $Li_x(C_5H_5N)_yFe_{2-z}Se_2$. Recent in-situ synchrotron x-ray total scattering study in the Li-PyH5-FeSe phase space, established such a metal site deficiency, which also relaxes the Fe-Fe network.[29] The enhanced disorder was ascribed to the highly reductive nature of the [Li-PyH5] intercalation medium used during the solvothermal synthesis of $Li_x(C_5H_5N)_yFe_{2-z}Se_2$[29,70], where excision of elemental species is reflected as a depleted Fe-site occupancy.

## C. First principles calculations

Our assessments, based on the analysis of the $Li_x(C_5H_5N)_yFe_{2-z}Se_2$ EXAFS, unveil the presence of Fe vacancies, with their concentration being related to the Li level. As such defects are known to sensitively affect the superconducting properties,[21,26,27] we invoke first principles density functional theory calculations to examine how atomic and electronic structures are modified in their presence. Since the EXAFS-based modeling of the local structure involved a single FeSe layer (SI §S1), the DFT calculations were also drawn on the same single-layer FeSe concept, approximating the isolated inorganic layers (d~ 16.5 Å) of the molecule intercalated system.

Calculations (SI S4) were carried out for the pristine $Fe_2Se_2$, assuming a unit cell (1×1) composed of two Fe and two Se atoms (Fig. 8a), and for $N \times N$ supercells, simulating the defected $Fe_{2-z}Se_2$ 2D layers, with variable Fe vacancy concentration, namely, the $Fe_{1.72}Se_2$ (3×3) and the $Fe_{1.92}Se_2$ (5×5) cases (Fig. 8b,c). Interestingly, the calculations suggest that the Fe-vacancy formation energies (SI §S4), at zero temperature, are relatively low, namely of $E_{Fe^*} = 0.402$ eV, for $Fe_{1.92}Se_2$ and 0.421 eV, for $Fe_{1.78}Se_2$, thus favoring their presence in single-layer β-FeSe. This behavior is in sharp contrast to the 2D transition metal dichalcogenides, where the vacancy formation energy for the metallic species is around ten times larger[71,72], thus raising the role of trigonal prismatic (M$Ch_6$) against pyramidal (M$Ch_4$) metal coordination, in strengthening the local bonding environments in dichalcogenides. Moreover, having relaxed both the atomic positions and the supercells of the 2D $Fe_{2-z}Se_2$ (z ≠ 0) layers, the calculations find that the square planar Fe-Fe network in the ab plane expands with increasing Fe vacancy concentration. This trend (Fig. 8d), is in strikingly good agreement with the relative modification of the in-plane Fe-Fe distances, with increasing Li-content, as derived by the EXAFS experiments on the expanded lattice $Li_x(C_5H_5N)_yFe_{2-z}Se_2$ system. Bearing in mind that strain changes significantly the electronic band structure (EBS) of 2D materials, such as the semiconducting transition metal di-chalcogenides[73], we examined how strain affects the EBS of single-layer $Fe_2Se_2$ (Fig. S6). The calculations show quite different modifications for a defect-free $Fe_2Se_2$ lattice stretched to the same degree of expansion as that generated by Fe-vacancies (*vide infra*).

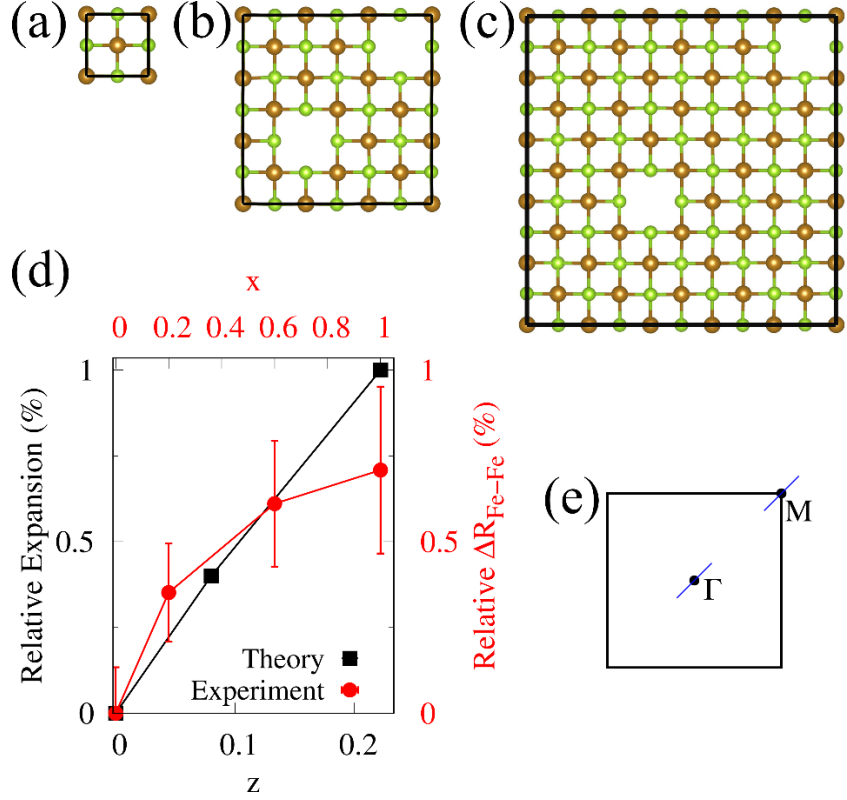

**Figure 8**. Ball and stick representation of the ab plane (top-view) for the DFT simulated single-layer iron selenide (Fe, brown and Se, green spheres) derivatives: (a) The primitive unit cell (1×1) of pristine $Fe_2Se_2$ and the supercells with Fe-vacancies, (b) a (3×3) lattice for $Fe_{1.72}Se_2$, and (c) a (5×5) lattice for $Fe_{1.92}Se_2$. (d) In-plane lattice expansion upon creation of Fe vacancies in single-layer iron selenides simulated by DFT (squares), and the variation of the EXAFS derived in-plane Fe-Fe distances, relative to the parent β-FeSe ($\Delta R_{Fe-Fe}$; circles), with increasing Li-content, in the $Li_x(C_5H_5N)_yFe_{2-z}Se_2$ series at 20 K. (e) Schematic of the first Brillouin zone of the pristine $Fe_2Se_2$ and the relevant paths (blue lines), close to the high-symmetry points, Γ and M.

For these reasons, we turn our focus on the band structure of the defected supercells. Calculations involved the pristine $Fe_2Se_2$ and the defected $Fe_{2-z}Se_2$ models, without taking into account interlayer interactions, i.e., assuming isolated single layers. The EBS of the pristine $Fe_2Se_2$ is in agreement with earlier works[74], with a hole pocket forming around the Γ point and an electron pocket emerging around the M point of the BZ (Fig. 9a-b). As the pristine monolayer exhibits energy states close to the $E_F$ around these points (Fig. S7), the vacancy bearing derivatives were assessed around the same regions. For the two cases studied by DFT (Fig. 9c-f), the overall effect of introducing Fe vacancies is two-fold: (a) the position of the electron and hole pockets relative to $E_F$ shifts and (b) additional bands are created above and below the hole pocket, with concomitant modification of the pristine monolayer bands (cf. Fig. 9b,d,f; emerging bands are described in SI §S4).

In particular, at the Γ point, the position of the hole pocket in $Fe_2Se_2$ shifts from above $E_F$ to an extreme of ~200 meV below the $E_F$ for the highly defected $Fe_{1.78}Se_2$ (Fig. 9f), and a moderate of ~50 meV for the weakly defected $Fe_{1.92}Se_2$ (Fig. 9d). The Fe-vacancy concentration in the latter is comparable to the Fe occupancy derived by EXAFS in $Li_{1.0}(C_5H_5N)_yFe_{2-z}Se_2$ (Table 1). Importantly, a substantial suppression of the hole pocket by ~80 meV below $E_F$ at the $\Gamma$ point has been experimentally observed in films of single-layer FeSe/SrTiO$_3$ ($T_c$~ 65 K)[75,76]. For the bulk $(Li_{1-x}Fe_x)OHFe_{1-y}Se$ ($T_c$ ~ 41 K) superconductor, with wide interlayer separation (d~ 9.3 Å), a related lowering of the hole pocket by ~50-70 meV is seen[77]. These experiments indicate that the suppression of the hole pocket near the $\Gamma$ point may be favored for high-$T_c$ 2D Fe-superconductors. A less systematic trend, though, is seen by DFT around the M point in $Fe_{2-z}Se_2$. Setting aside the extremely defected $Fe_{1.78}Se_2$ (Fig. 9e), DFT finds the bottom of the electron-like band also located below $E_F$ in the pristine $Fe_2Se_2$ (~80 meV) and the weakly defected $Fe_{1.92}Se_2$ (~50 meV) (Fig. 9a,c), following the trend seen by ARPES measurements[77] on related single-layer FeSe/SrTiO$_3$ (~60 meV) and bulk $(Li_{1-x}Fe_x)OHFe_{1-y}Se$ (~50 meV) systems.

Concerning superconductivity in such iron-selenides, high-$T_c$ has been associated with common band structure features, including the presence of electron pockets close to the BZ corners (M point) and the suppression of hole-like Fermi surface bands near the center ($\Gamma$ point) [75–77]. According to the present calculations, this is the case not only for the pristine and the electron doped single-layer (Fig. S7), but also for its Fe-vacancy derivatives $Fe_{2-z}Se_2$, resembling the low-dimensional inorganic layers in $Li_x(C_5H_5N)_yFe_{2-z}Se_2$. In view of these Fermi surface ingredients, FeSe sheet separation beyond the saturated value (d~ 8.6 Å)[14], when combined with electron donating spacers, may provide a view as of why $T_c$ is robust for the intercalated samples studied here.

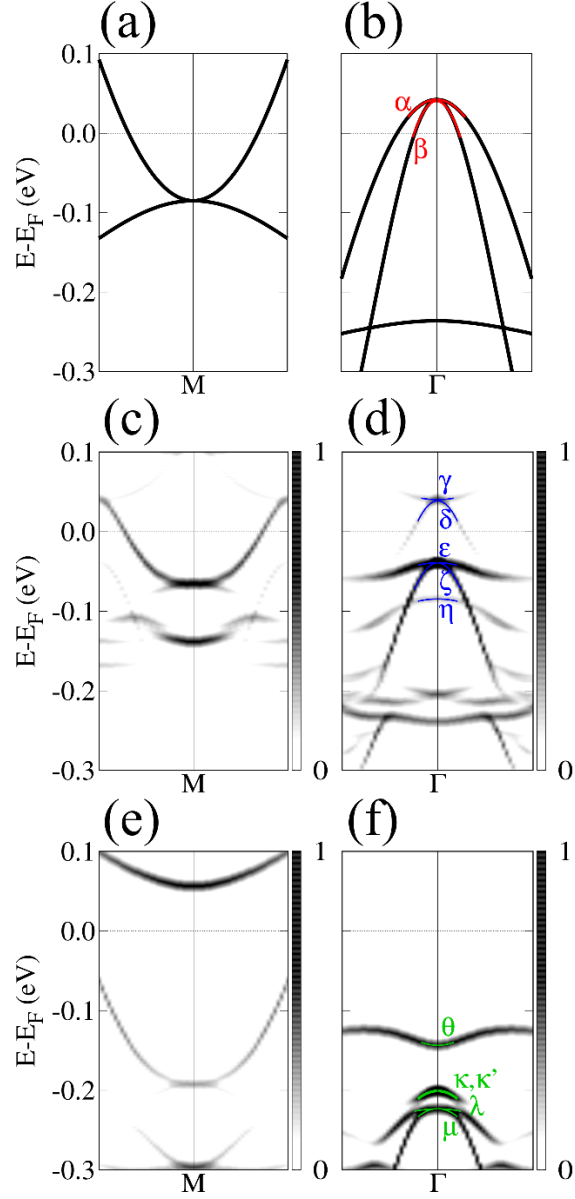

**Figure 9.** DFT calculations of the electronic band structure along the Γ and M paths (Fig. 8e) for the pristine $Fe_2Se_2$ monolayer (a, b), and the effective band structures of supercells with Fe-vacancies, namely, of $Fe_{1.92}Se_2$ (c, d) and $Fe_{1.78}Se_2$ (e, f). The lower-case Greek letters are guides for the evolution of the bands (see SI §S4) close to the Fermi level ($E_F$, horizontal dotted line) around the Γ point. The grey-scale bar shows the degree of projection from the supercell BZ to the primitive cell BZ (SI §S1).

## IV. CONCLUSIONS

XAS and XES were utilized in order to investigate the electronic and local structure correlations in the molecule intercalated iron selenide $Li_x(C_5H_5N)_yFe_{2-z}Se_2$ as a function of doping ($0 \leq x \leq 1$) and temperature. The XANES region validates the $Fe^{2+}$ ($Se^{2-}$) valence state of the materials studied here. The evolution of XANES as a function of Li-content ($x$), confirms doping-mediated local atomic rearrangement and progressive filling of electronic orbital states, with concomitant reduction of empty states near the Fermi level with respect to β-FeSe. The K$\beta$ emission spectra revealed that the raised $T_c$ intercalated compounds carry a low spin state, as well as a somewhat reduced Fe local magnetic moment compared to β-FeSe. Local structure assessments through EXAFS, reveal systematic increase of the intralayer bond lengths (cf. Fe-Se and Fe-Fe) with increased amount of doping. The nature of Fe-Se bond remains highly covalent and stiff, similar to the parent compound and consistent with that in related superconducting selenides. However, the Fe-Fe distance evolves to become significantly softer at high Li content, as conferred by the reduced lattice force constant of the Fe-Fe network. The local bond fluctuations infer larger in-plane configurational disorder arising from some Fe-site deficiency. As that, the relaxation of the Fe sub-lattice compensates for the anion height compression, giving rise to 'taller' than optimum Se to Fe-sheet normal distances ($h_z$). As high-$T_c$ (~44 K) is maintained, such conventional optimization measures become less relevant in intercalated systems where layers are spaced far away.

The strong FeSe slab isolation upon intercalation has been simulated by spin-polarized DFT calculations on single-layer pristine and defected $Fe_{2-z}Se_2$ derivatives ($z$, vacant sites). Calculations find that Fe vacancies are formed at a relatively low energy cost, an order of magnitude lower than that in 2D dichalcogenides, providing further support to the experimental evidence pointing to their presence. DFT suggests that the defected 2D lattice expands with increasing Fe vacancy concentration, following a trend that resembles that of the Fe-Fe network stretching, as derived by the EXAFS analysis. The electronic band structures around the high symmetry Γ and M points of the $Fe_{2-z}Se_2$ first BZ were calculated for the pristine ($z= 0$) isolated monolayer and the respective Fe-vacancy ($z\neq 0$) bearing supercells. High-$T_c$ in the Fe-site depleted model systems studied here has been associated with EBS features common to iron-selenides with widely-separated Fe-sheets (d > 8.6 Å). These include the presence of electron pockets close to the BZ corners (M point) and the suppression of hole-like Fermi surface bands near the center (Γ point), both supporting the enhanced $T_c$ and its robustness in the expanded lattice $Li_x(C_5H_5N)_yFe_{2-z}Se_2$. The work highlights that in materials with Fe-interlayer distance beyond the saturated value (d ~ 8.6 Å), parametrization of the $T_c$ is not a simple relation of the distortion of the basic $FeSe_4$ tetrahedral units (cf. bond lengths, Fe site vacancy), but a combined function of their interplay with the doping level generated by the electron-donating molecules and relevant modifications of the inorganic layer's electronic band structure.

**ACKNOWLEDGEMENTS**

This material is based upon research supported by the Office of Naval Research Global under award no. N62909-17-1-2126. The spectroscopy experiments were performed at BL22-CLÆSS beamline at ALBA Synchrotron (Barcelona, Spain) with the collaboration of the ALBA staff. Work at Brookhaven National Laboratory was supported by the U.S. Department of Energy, Office of Science, Office of Basic Energy Sciences (DOE-BES) under Contract no. DE-SC0012704. First principles calculations were performed with computational time granted from the National Infrastructures for Research and Technology S.A. (GRNET S.A.) in the National HPC facility ARIS, under project pa181005-NANOCOMPDESIGN.

**Supporting Information:** Additional methods for experiments and theory; details on EXAFS modeling; LeBail analysis; integrated absolute differences (IADs) from XES; XANES pre-edge features; EXAFS data, 20 and 300 K; bond angles and anion heights; supercell creation by DFT; vacancy formation by DFT; electronic band structures by DFT (DOCX).

**TABLE OF CONTENTS GRAPHIC**

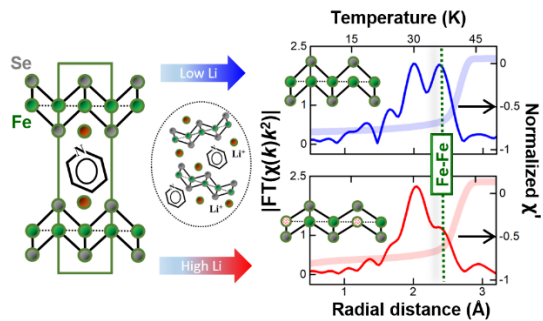

**TABLE OF CONTENTS SYNOPSIS**

Molecule intercalated iron selenides with widely separated layers, probed by x-ray absorption spectroscopy at the Fe K-edge, bear iron vacancies at high lithiation, but retain a robust superconducting state mediated by the materials' electronic structure.

# Li$_x$(C$_5$H$_5$N)$_y$Fe$_{2-z}$Se$_2$: a defect resilient expanded-lattice high-temperature superconductor

Alexandros Deltsidis,[a,b] Laura Simonelli,[c] Georgios Vailakis,[a,b] Izar Capel Berdiell,[a] Georgios Kopidakis,[a,b] Anna Krzton-Maziopa,[d] Emil S. Bozin,[e] and Alexandros Lappas[a,*]

[a]*Institute of Electronic Structure and Laser, Foundation for Research and Technology–Hellas, Vassilika Vouton, 71110 Heraklion, Greece*

[b]*Department of Materials Science and Technology, University of Crete, Voutes, 71003 Heraklion, Greece*

[c]*ALBA Synchrotron Light Source, Carrer de la Llum 2-26, 08290 Cerdanyola del Vallés, Spain*

[d]*Warsaw University of Technology, Faculty of Chemistry, Noakowskiego St. 3, 00-664, Warsaw, Poland*

[e]*Condensed Matter Physics and Materials Science Department, Brookhaven National Laboratory, Upton, NY 1197, USA*

[*] e-mail: lappas@iesl.forth.gr

# S1. Methods

**Experiments**

**Basic characterization.** All the procedures that involved sample preparations were carried out with the use of an Ar-circulating glovebox (MBRAUN, UNILab), with less than 1 ppm $O_2$ and $H_2O$, in order to avoid oxidation of the highly air-sensitive products. The phase purity of the samples was tested by powder X-Ray diffraction (PXRD) using a Bruker D8 Advance diffractometer utilizing monochromatic Cu-$K_a$ radiation 1.5418 Å (40 kV, 40 mA), equipped with a LYNXEYE 1D detector of active area 14.4 mm x 16 mm. The samples were studied by using a dedicated home-made, air-tight sample holder, designed for the study of the highly air-sensitive samples. Thermogravimetric (TG) measurements were performed with a TA Instruments SDT Q600 (v8.3 Build 101), simultaneous TGA-DSC system. Data were collected in the temperature range of 25 to 550 °C, with a rate of 10 °C/min, under Ar gas flow. AC susceptibility measurements were performed by utilizing an Oxford Instruments Maglab EXA 2000 vibrating sample magnetometer (VSM). AC susceptibility measurements were carried out at temperatures ranging from 5 to 50 K, with an AC field amplitude $H_{ac}$= 1 Oe, and a frequency f= 999 Hz.

**Powder X-ray diffraction.** LeBail full profile analysis of the PXRD for β-FeSe (Fig. S1a) was performed with the GSAS-II program suite[1]. The analysis confirmed the single-phase character of the starting material ($a = b = 3.7711(5)$ Å, $c = 5.5230(6)$ Å, P4/nmm) that was required for the solvothermal synthesis of pure $Li_x(C_5H_5N)_yFe_{2-z}Se_2$ intercalants. Although the limited 2θ range (due to the air-tight sample holder) for the latter, did not permit further LeBail analysis, the Bragg reflections of the intercalated phase were indexed on the basis of the $ThCr_2Si_2$[2] (122) structural type (Fig. 1).[3] The intercalants adopt an interlayer distance d = 16.2 Å that is characterized by the lowest-angle (00*l*) reflection at $2θ \cong 5.5°$. The (*hkl*)* indices, denote Bragg reflections belonging to the PbFCl[4] (11) structural type of the parent β-FeSe material (d = 5.5 Å). For the x = 0.2 sample, PXRD suggests some remnant β-FeSe component that coexists with the main expanded-lattice intercalated phase.

**Magnetic measurements.** The AC susceptibility versus temperature for the all the intercalated samples studied is compared against the β-FeSe in Fig. 1b,c. The volume susceptibilities have been normalized so that $4πχ$ = -1 (SI units). For the low doping level (x = 0.2), the two-phase behavior (already established by PXRD) is marked by the weak 8 K superconducting transition due to the remnant β-FeSe phase and the strong intrinsic superconducting transition at 44 K, attributable to the intercalated phase (Fig. 1b). For the other two doping levels (x = 0.6 and x = 1.0), the 44 K high-$T_c$ transition due to the intercalated phase is detected together with an increased paramagnetic signal above the onset of $T_c$. This second contribution may be understood when the strongly reducing character of the [Li-$C_5H_5N$] intercalation medium is considered[5]. The highly reductive environment of the solution may lead to excision of elemental species (e.g. Fe) from the β-FeSe matrix, which although not traceable by PXRD, are detected by AC susceptibility. Moreover, the imaginary part of the magnetic susceptibility for the intercalated samples is governed by intense peaks, appearing at ~40 K (Fig. 1c), which are attributed to the coupling losses between the superconducting grains of the samples.

**Thermogravimetric (TG) measurements.** TG measurements were obtained for the as-prepared $Li_x(C_5H_5N)_yFe_{2-z}Se_2$ (x = 0.2, 0.6, 1.0) samples (Fig. S1b). The TG curve for the low doping level (x= 0.2) shows a single mass drop ($\Delta w_1$ ~14%) below $T_1 \cong 140$ °C, while for the other two doping levels (x = 0.6, 1.0) there is a second mass loss step, $\Delta w_2$, below about $T_2 \cong 250$ °C (Fig. S1b; Table S1). The relatively lower temperature of $\Delta w_1$ for the x= 0.2 and the presence of non-intercalated β-FeSe phase in this sample (Fig. 1, main text), suggest that under the chosen conditions of the solvothermal reaction, $C_5H_5N$ molecules may be more weekly bound, such as when co-intercalated locally, as at the edge of β-FeSe crystallites. However, the higher temperature weight loss of $\Delta w_2$, at T< $T_2$, grows at the expense of the lower-temperature $\Delta w_1$ as the Li content increases from x= 0.6 to 0.1, in good agreement with the single-phase character of these intercalated samples (Fig. 1, main text). The behavior below 250 °C is analogous to that in related, expanded-lattice $Li_x(C_6H_{16}N_2)_yFe_{2-z}Se_2$ [6,7] and $Li_x(C_8H_{11}N)_yFe_{2-z}Se_2$ [8] derivatives, where the mass loss was assigned to the de-intercalation (or desorption) of the respective organic molecules. We estimated the y-content (Table S1) in the chemical formula of $Li_x(C_5H_5N)_yFe_{2-z}Se_2$ by assigning the high-temperature mass loss (Fig. S1b) exclusively to the de-intercalation of the organic molecules ($C_5H_5N$), while the remaining mass, above 250 °C, was allocated to the inorganic residuals (Li + FeSe).

**Table S1.** TG curves analysis and an estimate of the y-content in $Li_x(C_5H_5N)_yFe_{2-z}Se_2$.

| Li nominal, [x] | $T_1$ (°C) | $T_2$ (°C) | $\Delta w_1$ (%) | $\Delta w_2$ (%) | $C_5H_5N$, [y] |
|---|---|---|---|---|---|
| 0.2 | 141 | N/A | 14.3 | N/A | 0.30 |
| 0.6 | 122 | 241 | 11.4 | 9.5 | 0.49 |
| 1.0 | 122 | 246 | 7.7 | 11.9 | 0.45 |

**Core-level absorption spectroscopy.** The background subtracted Fe and Se K-edge XANES spectra have been normalized with respect to the atomic absorption estimated by a quadratic polynomial fit in the post-edge region, within the Athena software[9]. A standard procedure based on polynomial spline function fit to the pre-edge subtracted spectra was used to extract the EXAFS oscillations[10]. While the XANES provides information on the electronic state of the material, the EXAFS region is dominated by oscillations, which reflect the local structure around the absorber. In order to quantify the local structure parameters the $k^2$-weighted EXAFS oscillations have been modelled on the basis of the single scattering approximation[11], by the standard equation:

$$k^n \chi(k) = S_0^2 \sum_i N_i F_i(k_i) k_i^{n-1} e^{-2\sigma_i^2 k_i^2} e^{-\frac{2r_i}{\lambda_i(k_i)}} \times \frac{\sin[2k_i r_i + \varphi_i(k_i)]}{r_i^2} \quad (1)$$

where $S_0^2$ is the EXAFS reduction factor due to many-body effects. $N_i$ is the number of neighbouring atoms at a distance $r_i$, $F_i(k_i)$ is the backscattering amplitude, $k_i$ is the wave number of the photoelectron, $\sigma_i^2$ is the EXAFS Debye-Waller factor measuring the mean square relative displacements (MSRD) of the photoabsorber-backscatterer pairs due to their thermal motion and $\lambda_i$ is the photoelectron mean free path. The $\varphi_i$ is the phase shift implemented in the backscattered wave. The EXAFS modelling is carried out using the Artemis software[9] that uses FEFF8 code[12]. During modeling, the magnitude of $S_0^2$ has been fixed at a constant value after having analyzed different EXAFS scans, over a broad temperature range and then by taking the average of $S_0^2$ as derived from these scans. A similar procedure has been followed for obtaining the energy threshold (E₀).

**EXAFS modeling.** The modeling of the EXAFS data is pursued in the context of single scattering approximation (Eq. 1). The starting crystal model utilized in the analysis discussed here rests on the simple tetragonal (P4/nmm) structure of the layered β-FeSe[13]. Basically, the local structure is approximated by a single FeSe layer, where the first coordination shell around Fe (Se) involves 4 Se (Fe) atoms at a distance ~2.39 Å. Assuming constant number of atoms surrounding the absorber, fitting the data with a single shell requires refining two parameters, namely, the Fe-Se distance and their corresponding thermal parameter ($\sigma_i^2$). For the Fe K-edge spectra, a second coordination shell is considered, composed of 4 Fe atoms at a distance ~2.67 Å. So, in the case of a two-shell, always assuming the number of neighboring atoms fixed, 4 parameters in total are free to be refined, i.e., Fe-Se and Fe-Fe distances and the corresponding two Debye-Waller factors, $\sigma_i^2$. (cf. Mean Square Relative Displacements – MSRDs). The MSRDs for a pair of atoms reflect the lattice configurational disorder expressed through the sum of temperature-independent ($\sigma_0^2$; static) and temperature-dependent ($\sigma_d^2$; dynamic) terms, $\sigma^2 = \sigma_0^2 + \sigma_d^2(T)$ [14]. The temperature dependent MSRDs for the Fe-Fe and the Fe-Se pairs, as a function of temperature and doping, have been extracted during the fitting of a local lattice model to the EXAFS experiments. In the context of the harmonic and single scattering approximation, the MSRDs can be described by the correlated Einstein model[14,15], where the temperature dependent term is given:

$$\sigma_d^2(T) = \frac{\hbar^2}{2\mu k_B \theta_E} \coth\left(\frac{\theta_E}{2T}\right) \quad (2)$$

where $\mu$ is the reduced mass of the respective atom pair and $\theta_E$ is the Einstein temperature, related to the Einstein frequency ($\omega_E = k_B \theta_E/\hbar$). During the modelling of the EXAFS data, the maximum number of independent fitted parameters defined by $2\Delta k\Delta R/\pi$ (~ 9), has been always kept above the number of refined parameters.

In order to assess the occupational disorder within the FeSe slabs, we also performed least square refinements of the coordination number (N; nearest-neighbor Fe atoms around the Se absorbing site) in the afore-mentioned local lattice model assuming a single scattering path in the Se K-edge. The analysis suggests that in the low Li (x= 0, 0.2, 0.6) compositions of Li$_x$(C$_5$H$_5$N)$_y$Fe$_{2-z}$Se$_2$, Se is coordinated by N~ 4 neighboring Fe atoms, inferring a near stoichiometry (Table 1 – main text). However, at elevated Li content (x = 1.0) the refined coordination was N~ 3.88(9), proposing site deficiency and the likely presence of Fe vacancies.

**Theoretical calculations**

**Band unfolding.** In order to study the band structure of a single-layer iron selenide without and with Fe vacancies, a 1× 1 unit cell of $Fe_2Se_2$ (pristine) and $N \times N$ supercells of $Fe_{2-z}Se_2$ are used, respectively. Supercells of different sizes are employed in order to study different Fe vacancy concentration. The corresponding effective band structures are obtained by unfolding from the supercell to the pristine $Fe_2Se_2$ unit cell Brillouin zone[16]. Band structures of simulated cells with different size, e.g., Fig. S5a-c, need careful consideration since the points change position and periodicity in the different Brillouin zones (Fig. S5d). Standard DFT electronic structure calculations with the simulated cells of Fig. S5a-c produce the band structures shown in Fig. S5e-j. The electronic band structure of the pristine $Fe_2Se_2$ unit cell (Fig. S5a) is shown in Fig. S5e,f. The band structures of $Fe_{1.72}Se_2$ (Fig. S5b) and $Fe_{1.92}Se_2$ (Fig. S5c) are shown in Fig. S5g,h and Fig. S5i,j, respectively. Our goal is to compare the band structures of the different supercells and observe the modification of the bands with the introduction of the Fe vacancies. The effective band structure (EBS) is calculated by projecting the eigenstates of the supercells on a different Brillouin zone, in our case the 1$^{st}$ Brillouin zone of the pristine FeSe (Fig. 8d or Fig. S5d). The effective band structure is calculated through the spectral function:

$$A(\mathbf{k}, E) = \sum_{v} P_{K,v}(\mathbf{k}) \delta(E_{K,v} - E) \tag{3}$$

where $\mathbf{k}$ is a point at the targeted 1$^{st}$ Brillouin zone, $E$ is the energy and $E_{K,v}$ is the eigenvalue of the state $K, v$ with $K$ a point of the 1$^{st}$ Brillouin zone of the simulation unit cell and $v$ the quantum number. The projection $P_{K,v}(\mathbf{k})$ is defined as $P_{K,v}(\mathbf{k}) = \sum_g |C_{K,v}(\mathbf{g} + \mathbf{k})|^2$ (the value of the grey-scale bar in Fig. 9), where $C_{K,v}(\mathbf{g} + \mathbf{k})$ are the coefficients on the plane wave expansion of the eigenstate $K, v$ and $\mathbf{g} = n_1 \mathbf{b_1} + n_2 \mathbf{b_2}$ with $n_{1,2}$ are integer numbers and $\mathbf{b_{1,2}}$ the reciprocal lattice vectors of the targeted 1$^{st}$ Brillouin zone.

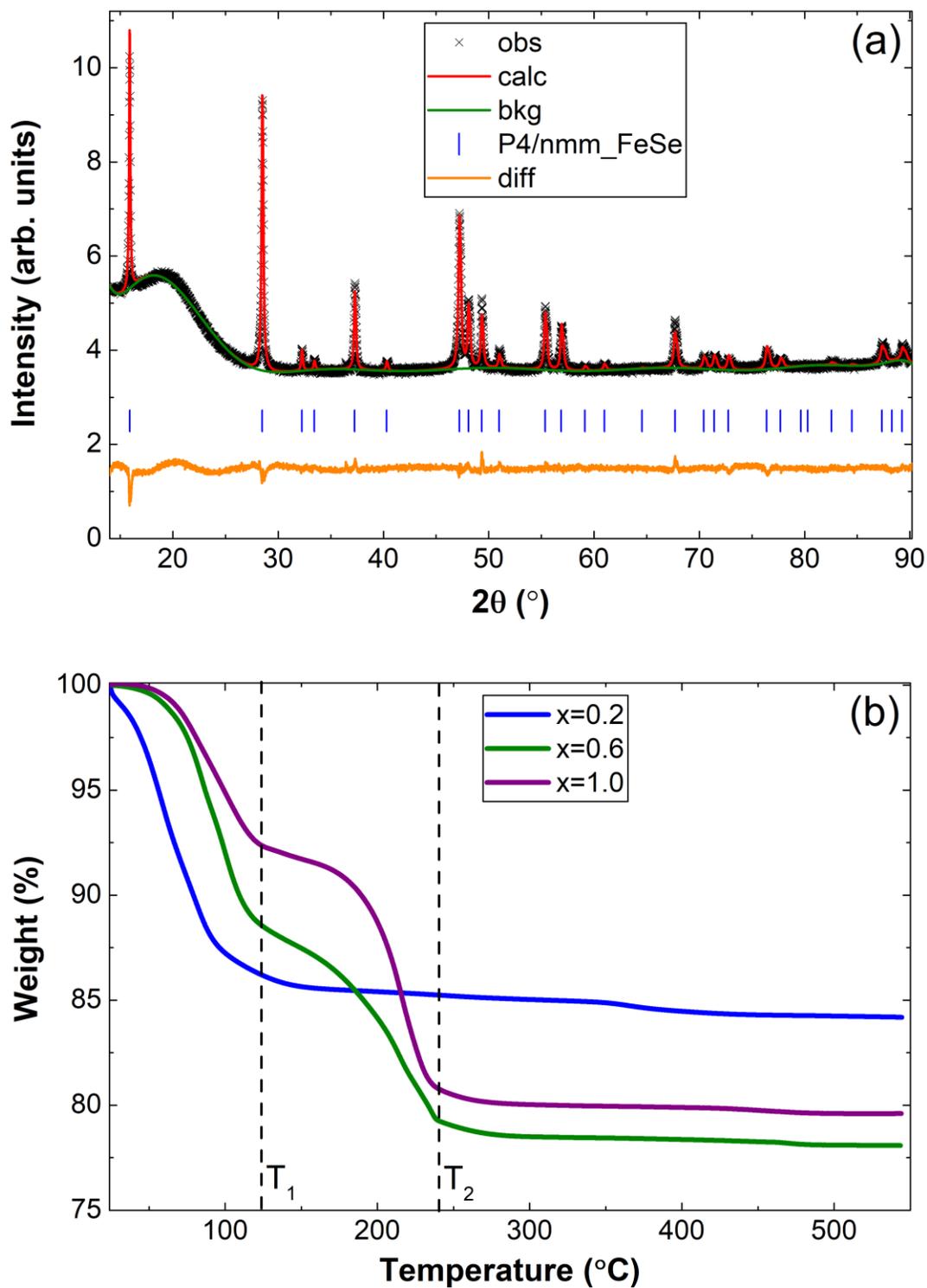

**Figure S1.** (a) LeBail full profile analysis of the parent material (β-FeSe), indexed on the basis of the PbFCl structure type (P4/nmm symmetry); $a = b = 3.7711(5)$ Å, $c = 5.5230(6)$ Å. (b) TG, weight loss curves of $Li_x(C_5H_5N)_yFe_{2-z}Se_2$ (x = 0.2, 0.6 and 1.0) samples, heated under flowing Ar gas; $T_1$ and $T_2$ mark characteristic temperatures for the de-intercalation of the $C_5H_5N$ molecules.

## S2. Electronic structure insights

**Absorption spectra.** The pre-edge peak (#A) of the Fe K-edge has been modelled with a Gaussian fit in order to quantify the intensity reduction from parent to intercalated materials. Specifically, a spline function is extrapolated in the energy region [7108-7116] eV, in order to remove the background, enabling the modelling of the pre-edge peak (#A). The reduced mixing between the Fe $3d$ and $4p$ orbitals is mirrored in the intensity reduction which for the parent β-FeSe material is 0.08(2), in contrast with the expanded lattice systems where it is found to be ~0.07(2).

**X-ray emission spectra.** A Fe K$\beta$ emission spectrum can provide information on the local Fe moment[17–19]. Indeed, the integrated absolute value (IAD) of the spectra with respect to a reference is proportional to the Fe local magnetic moment. The procedure followed for the extraction of the IAD involves (i) normalization of the spectrum area, (ii) subtraction of a reference spectrum from all the spectra and (iii) integration of the absolute values of the respective difference spectra. Here, we have subtracted the room temperature spectra of the parent material (β-FeSe) from the respective spectra of each intercalated material $Li_x(C_5H_5N)_yFe_{2-z}Se_2$ ($0.2 \leq x \leq 1.0$). The relative variation of the IAD value is calculated in the energy range 7028-7078 eV.

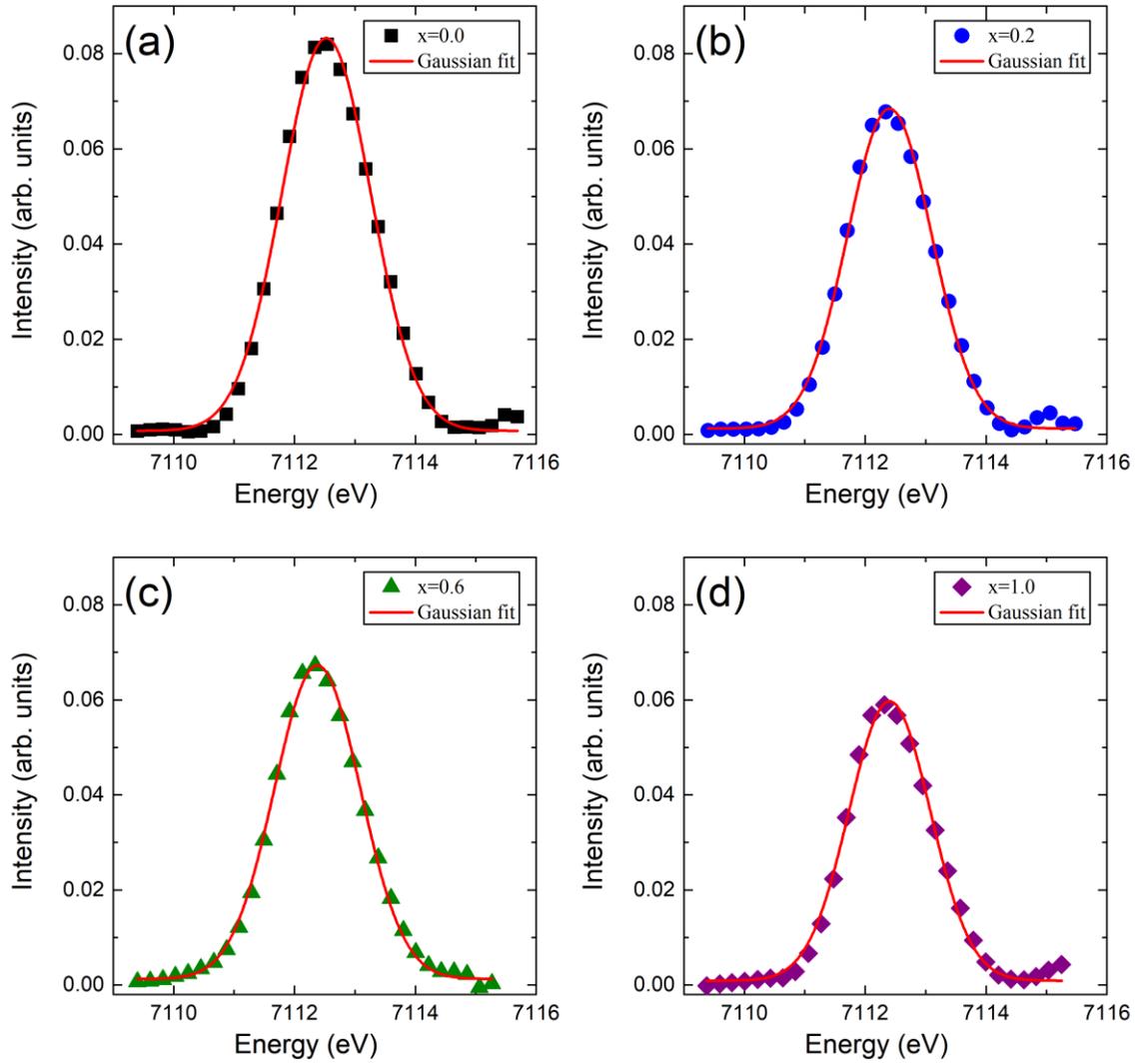

**Figure S2.** Gaussian fit of the pre-edge peak (#A) after background subtraction for the (a) parent compound β-FeSe (b) $Li_{0.2}(C_5H_5N)_yFe_{2-z}Se_2$ (c) $Li_{0.6}(C_5H_5N)_yFe_{2-z}Se_2$ and (d) $Li_{1.0}(C_5H_5N)_yFe_{2-z}Se_2$ (T= 20 K).

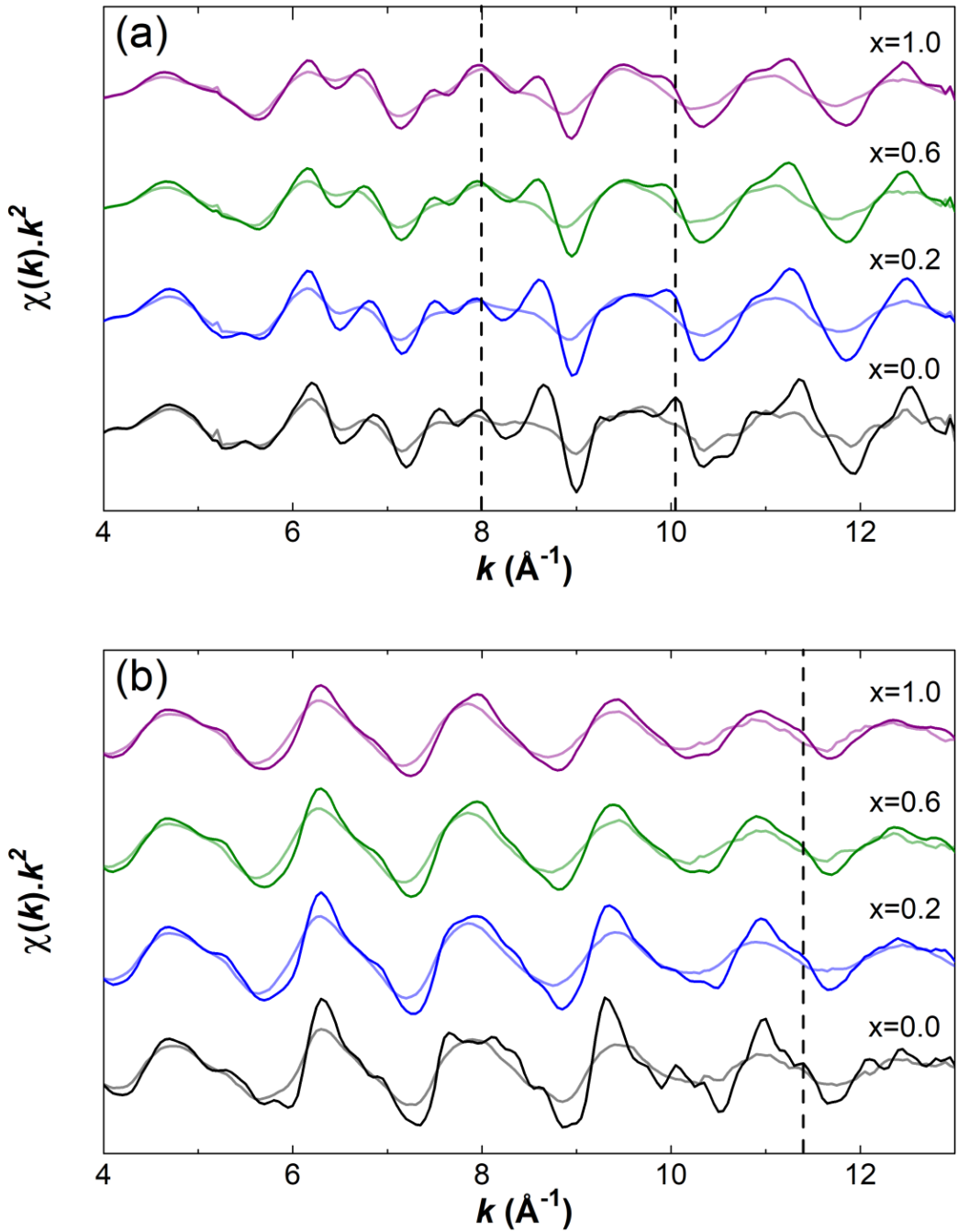

**Figure S3.** The EXAFS oscillations of the Li$_x$(C$_5$H$_5$N)$_y$Fe$_{2-z}$Se$_2$ ($0 \leq x \leq 1$) samples at 20 K and 300 K (light colored lines) extracted from the x-ray absorption spectra of (a) Fe and (b) Se K-edge, respectively. Here, we show the EXAFS signal corresponding to Li-contents of x= 0.0 (black), x= 0.2 (blue), x= 0.6 (green) and x= 1.0 (purple). The data are shifted vertically for clarity. Signatures of subtle structural changes are marked by vertical dashed lines, either at the Fe K-edge (a), or at the Se K-edge at k~ 11.5 Å$^{-1}$ (b).

## S3. Local structure insights

**Intercalant driven local distortions.** The anion height, i.e., the Se atom normal distance from the Fe-Fe plane and the associated Se-Fe-Se bond angle have been proposed to be key parameters for superconductivity in iron-based materials, having a direct relationship with the $T_c$ [20,21]. The measured bond lengths from EXAFS permit to determine these parameters. Assuming tetrahedral coordination of Se we can calculate the respective anion height ($h_z$) and bond angle ($\alpha$) using the following equations:

$$\alpha = \pi - 2\cos^{-1}\left(\frac{R_{Fe-Fe}}{\sqrt{2}R_{Fe-Se}}\right) \quad (4)$$

$$h_z = \sqrt{R_{Fe-Se}^2 - \frac{1}{2}R_{Fe-Fe}^2} \quad (5)$$

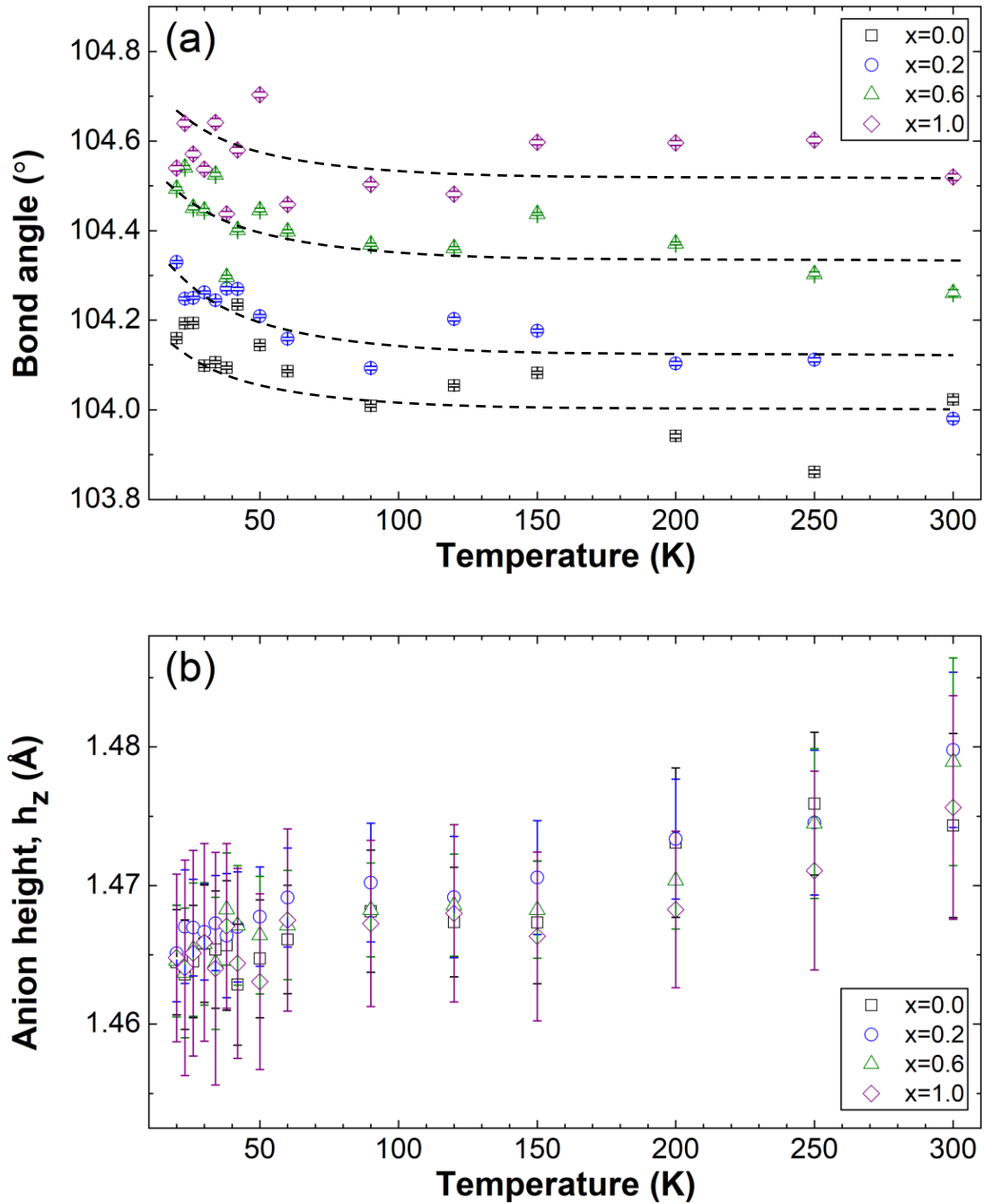

**Figure S4.** The temperature dependence of (a) the bond angle α and (b) the anion height, $h_z$, determined from the intra-layer bond distances ($R_{Fe-Se}$, $R_{Fe-Fe}$) extracted from fitting the measured Fe K-edge EXAFS spectra of the $Li_x(C_5H_5N)_yFe_{2-z}Se_2$ series. The dashed lines are a guide to the eye.

## S4. First principles calculations

The unit cell for our calculations contains two Fe and two Se atoms. For the single layer FeSe, shown in Fig. S5a (Fig. 8a), we find that the ground state of the pristine material is antiferromagnetic (AF) with lattice constant $a = 3.708$ Å, which is close to the 3.771 Å found in this work from the EXAFS analysis of the parent material. This unit cell is repeated $N$ times in each planar direction to create $N \times N$ supercells in order to study defected systems. We present two cases with different concentrations of Fe vacancies. These two systems with vacancies consist of supercells $3 \times 3$ and $5 \times 5$ with two Fe vacancies in each case, so that the system remains AF, corresponding to $Fe_{1.78}Se_2$ and $Fe_{1.92}Se_2$ (Fig. S5b,c and Fig. 8b,c), respectively. These supercells produce a tetragonal periodicity for the vacancy, with length period $S_{N \times N} = N \frac{\sqrt{2}}{2} a$, where $N = 3, 5$, respectively. The concentration of vacancies in the $Fe_{1.92}Se_2$ $5 \times 5$ supercell, is comparable to the highly lithiated sample (x= 1.0) studied by EXAFS (Table 1).

**Vacancy formation energy.** The formation energy of the metallic species is calculated at zero temperature in the two supercells ($3 \times 3$ and $5 \times 5$), representing the two defected systems with $Fe_{1.78}Se_2$ and $Fe_{1.92}Se_2$, respectively. The Fe-vacancy formation energy is defined as:

$$E^*_{Fe} = \frac{E_{N \times N} + n\,\mu_{Fe} - N^2\,E_{pr}}{n} \tag{6}$$

where $E_{N \times N}$ is the energy of the $N \times N$ supercell with $n$ Fe vacancies, $\mu_{Fe}$ is the chemical potential of Fe atoms calculated from the bulk BCC ferromagnetic state of Fe, and $E_{pr}$ is the energy of the pristine $Fe_2Se_2$ monolayer.

**Vacancy mediated emerging bands.** Focusing on EBS plots around the Γ points (main text, Fig. 9b,d,f) the effects of Fe-vacancies become quite pronounced. In the pristine iron selenide, the hole pocket consists of two bands, denoted as α and β (Fig. 9b). For z=0.08, these bands, denoted as ε and ζ (Fig. 9d) change significantly. It can be seen that the introduction of Fe vacancies affects the hole pocket by flattening the α band to the ε band. Addition ally, three new bands appear, namely, two above E$_F$ (named, γ and δ), and one below the hole pocket (η). Comparing with z=0.22 (Fig. 9f), there is an apparent change of the vacancy band from γ to θ and from η to λ. The hole pocket bands ε, ζ, together with the vacancy band δ, cannot be distinguished now between κ, κ´ and μ. Interestingly, the curvature of the top of the hole pocket (band κ) is similar to the pristine (band α), even though at z=0.08 the hole pocket band is flattened (band ε).

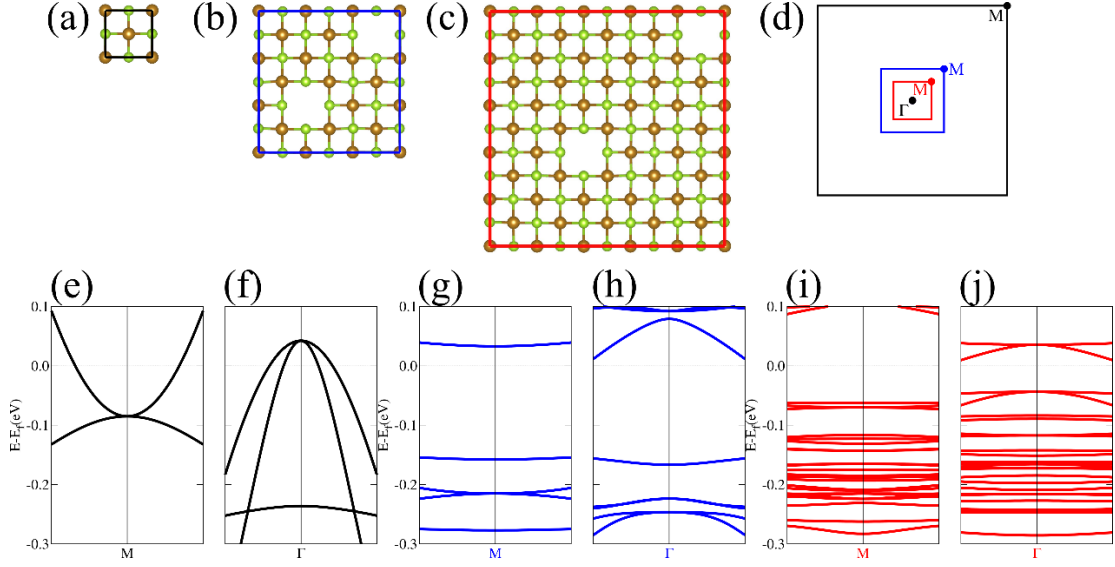

**Figure S5.** The simulated unit cells of $Fe_2Se_2$ (a), $Fe_{1.72}Se_2$ (b) and $Fe_{1.92}Se_2$ (c). The first Brillouin zones (BZ) and high-symmetry points of relevance (M and $\Gamma$), for all three unit cells (d), namely, the pristine $Fe_2Se_2$ (black), as well as the defected $Fe_{1.72}Se_2$ (blue) and $Fe_{1.92}Se_2$. (red). The corresponding band structures close to M and $\Gamma$ points for the BZ of each simulated unit cell, namely of $Fe_2Se_2$ (e, f), $Fe_{1.72}Se_2$ (g, h) and $Fe_{1.92}Se_2$ (i, j), suggesting that band structure unfolding may be necessary to uncover hidden features of such diagrams (g to j).

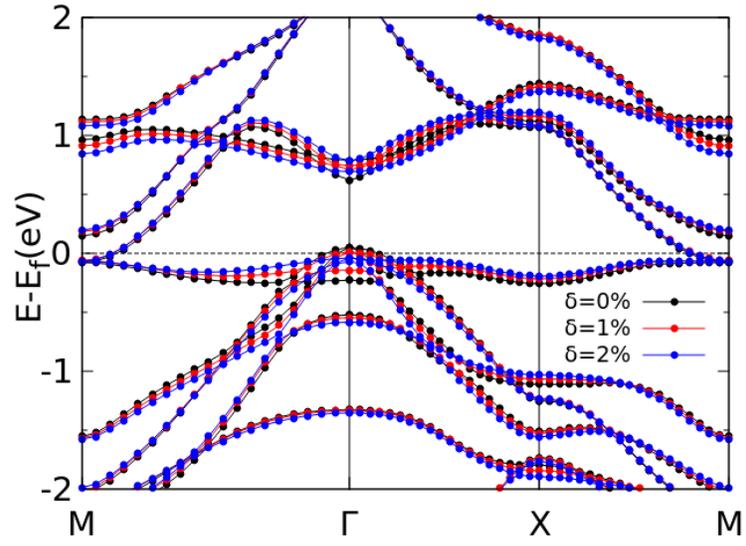

**Figure S6.** The band structure of strained single-layer $Fe_2Se_2$. Black, red, blue points correspond to unit cells uniformly stretched by 0%, 1%, 2%, respectively.

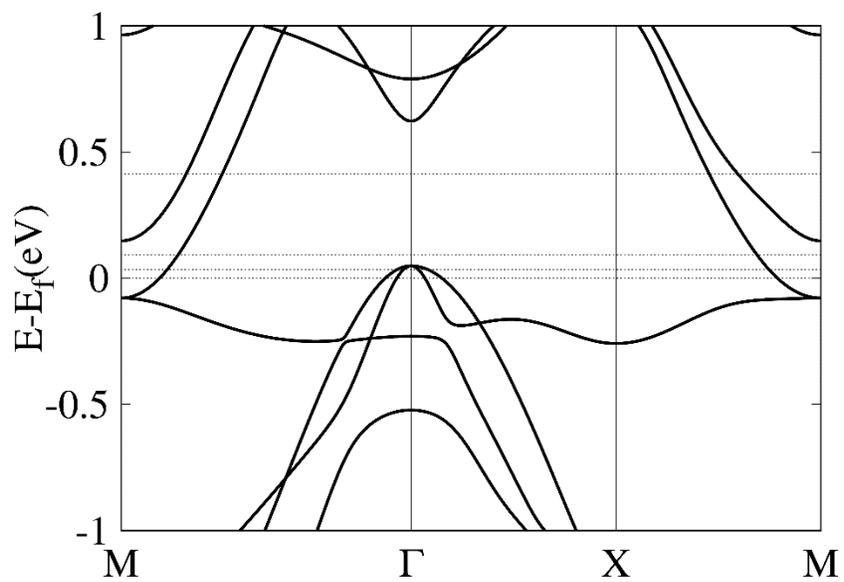

**Figure S7.** The band structure of the pristine single-layer $Fe_2Se_2$. The dashed lines represent the Fermi levels ($E_F$) for different values of excess electrons per Fe atom, from bottom up at 0.0, 0.1, 0.2 and 0.5 e/Fe.